\documentclass[traditabstract]{aa} % for the abstract without structuration 
\usepackage{graphicx}
\usepackage[latin1]{inputenc}
\usepackage{amsmath}
\usepackage{amsfonts}
\usepackage{amssymb}
\usepackage{natbib}
\usepackage{txfonts}
\usepackage{enumerate}
\usepackage{morefloats}
\usepackage{subfig}
\newcommand{\Daophot}{D\textsc{ao}P\textsc{hot}II{}}
\newcommand{\RA}[3]{$\alpha = #1^{\mathrm{h}}#2^{\mathrm{m}}#3^{\mathrm{s}}$}
\newcommand{\Dec}[3]{$\delta = #1\degr #2\arcmin #3\arcsec$}
\newcommand{\Geo}[3]{$#1\degr #2\arcmin #3\arcsec $}

\newcommand{\Msun}{\ensuremath{\,{\rm M}_\odot}}               % Solar mass symbol
\newcommand{\Rsun}{\ensuremath{\,{\rm R}_\odot}}                  % Solar radius symbol
\newcommand{\Mjup}{\ensuremath{\,{\rm M}_{\rm Jup}}}              % Jupiter mass symbol
\newcommand{\Rjup}{\ensuremath{\,{\rm R}_{\rm Jup}}}              % Jupiter radius symbol
\newcommand{\tci}{{TCI }}%two-colour instrument}
\newcommand{\tcI}{{TCI}}%two-colour instrument}
\newcommand{\elec}{\mathrm{e^-}}

\graphicspath{{figs/}}

\begin{document}

   \title{The Two-Colour EMCCD Instrument for the Danish 1.54m Telescope and SONG\thanks{Based on data collected with the Danish 1.54m telescope at ESO's La Silla Observatory.}}

%    \subtitle{Following the design of the SONG system}

   \author{Jesper~Skottfelt   \thanks{email:skottfelt@nbi.dk} \inst{\ref{nbi},\ref{starplan}}
 	\and D.~M.~Bramich \inst{\ref{qatar}}
 	\and M.~Hundertmark \inst{\ref{nbi}}
	\and U.~G.~J{\o}rgensen \inst{\ref{nbi},\ref{starplan}}
 	\and N.~Michaelsen \inst{\ref{nbi}}
	\and P.~Kj{\ae}rgaard \inst{\ref{nbi}}
 	\and J.~Southworth \inst{\ref{keele}}
 	\and A.~N.~S{\o}rensen \inst{\ref{nbi}}
	\and M.~F.~Andersen \inst{\ref{aarhus}}
	\and M.~I.~Andersen \inst{\ref{nbi}} 
	\and J.~Christensen-Dalsgaard \inst{\ref{aarhus}}
	\and S.~Frandsen \inst{\ref{aarhus}}
	\and F.~Grundahl \inst{\ref{aarhus}}
	\and K.~B.~W.~Harps{\o}e \inst{\ref{nbi},\ref{starplan}}
	\and H.~Kjeldsen \inst{\ref{aarhus}}
	\and P.~L.~Pall\'e \inst{\ref{canary},\ref{tenerife}}
%	\and A.~Trivi\~{n}o~Hage \inst{\ref{canary},\ref{tenerife}}
     }

   \institute{Niels Bohr Institute, University of Copenhagen, Juliane Maries Vej 30, 2100 K\o{}benhavn \O{}, Denmark \label{nbi}
	\and Centre for Star and Planet Formation, Natural History Museum, University of Copenhagen, \O{}stervoldgade 5-7, 1350 K\o{}benhavn K, Denmark \label{starplan}
    \and Qatar Environment and Energy Research Institute, Qatar Foundation, P.O. Box 5825, Doha, Qatar  
\label{qatar} 
%    \and Institut d'Astrophysique et de G\'eophysique, Universit\'e de Li\`ege, All\'ee du 6 Ao\^ut, B\^at. B5c, 4000 Li\`ege, Belgium \label{belgium}
	\and Astrophysics Group, Keele University, Staffordshire, ST5 5BG, UK \label{keele}
	\and Stellar Astrophysics Centre, Department of Physics and Astronomy, Aarhus University, Ny Munkegade 120, DK-8000 Aarhus C, Denmark \label{aarhus}
	\and Instituto de Astrof\'{i}sica de Canarias (IAC), E-38200 La Laguna, Tenerife, Spain \label{canary}
	\and Dept. Astrof\'{i}sica, Universidad de La Laguna (ULL), E-38206 Tenerife, Spain \label{tenerife}  
    }

   \date{Submitted 1 Nov 2014 / \today}

% \abstract{}{}{}{}{} 
% 5 {} token are mandatory
 
  \abstract
  {We report on the implemented design of a two-colour instrument based on electron multiplying CCD (EMCCD) detectors. This instrument is currently installed at the Danish 1.54m telescope at ESO's La Silla Observatory in Chile, and will be available at the SONG (Stellar Observations Network Group) 1m telescope node at Tenerife and at other SONG nodes as well.
 We present the software system for controlling the two-colour instrument and calibrating the high frame-rate imaging data delivered by the EMCCD cameras.
 An analysis of the performance of the Two-Colour Instrument at the Danish telescope shows an improvement in spatial resolution of up to a factor of two when doing shift-and-add compared with conventional imaging, and that it is possible to do high-precision photometry of EMCCD data in crowded fields.
 The Danish telescope, which was commissioned in 1979, is limited by a triangular coma at spatial resolutions below $0\farcs5$ and better results will thus be achieved at the near diffraction limited optical system on the SONG telescopes, where spatial resolutions close to $0\farcs2$ have been achieved. 
 Regular EMCCD operations have been running at the Danish telescope for several years and have produced a number of scientific discoveries, including microlensing detected exoplanets, the detection of previously unknown variable stars in dense globular clusters and the discovery of two rings around the small asteroid-like object (10199) Chariklo.  }

  % context heading (optional)
%{}
%  % aims heading (mandatory)
%{A } 
%  % methods heading (mandatory)
%{B } %   
%  % results heading (mandatory)
%{C } %   
%  % conclusions heading (optional), leave it empty if necessary 
%{D }


   \authorrunning{J.Skottfelt et al.}
   \titlerunning{Two-Colour EMCCD Instrument}
   
   \maketitle
%
%________________________________________________________________

\section{Introduction}
For most ground-based optical and near-infrared (NIR) telescopes today, seeing is the limiting factor in spatial resolution. 
Turbulence in the layers of the atmosphere distorts the wavefront of light coming from astronomical objects and effectively smears out the signal over a larger area in the focal plane, called the seeing disc.
In order to mitigate the effects of the atmospheric distortion, observatories are therefore often placed at sites located several kilometres above sea level.
However, even at the best sites the seeing rarely falls below $0\farcs5$, measured in the full width at half maximum (FWHM) of the point spread function (PSF), and a seeing below $1\arcsec$ is considered to be good. 
The theoretical limit for the spatial resolution of a telescope is known as the diffraction limit and it can be calculated as $1.22\lambda/D$, where $\lambda$ is the wavelength and $D$ is the telescope diameter. 
A seeing of $1\arcsec$ thus corresponds to the diffraction limit of a 10\,cm telescope at 500\,nm. 

Adaptive optics (AO) systems can be used for correcting the wavefront, by performing a real-time analysis of a bright nearby star, or a laser guide star.
AO systems on large telescopes ($>$5m) have been shown to reach spatial resolutions down to a few tens of milliarcseconds (mas), but they are very complicated and expensive systems and they are therefore not cost-effective for smaller telescopes. 

An alternative method for improving the spatial resolution is high frame-rate imaging. 
%At short enough exposure times (100\,ms or less) it is possible to make snapshots of the atmosphere as it changes. 
At very short exposure times ($\sim 10\,\text{ms}$) it is possible to make snapshots of the wavefront before it is changed to a new configuration by the atmospheric turbulence.
By analysing the speckle pattern in each snapshot, one may counteract the atmospheric disturbances by stacking the brightest speckles on top of each other and thus improve the spatial resolution.
This technique, known as speckle imaging (or just shift-and-add), is based on the work by \citet{Fried1966}.
Even using longer exposure times ($\sim 100\,\text{ms}$) it is possible to mitigate the effect of image motion created by atmospheric turbulence, by doing shift-and-add.

\citet{Fried1978} showed that there is some probability of obtaining a lucky short exposure, i.e. an exposure where the wavefront has passed almost unperturbed through the atmosphere.
If only these few lucky frames are used, then it is possible to achieve very high spatial resolution; a technique known as Lucky Imaging \textbf{(LI)}. 
However, as this probability is inversely proportional to the telescope diameter, only very few lucky frames can be obtained on large telescopes. 

Charge-couple devices (CCDs) are unsurpassed for visual and NIR observations as they provide spatial information, high quantum efficiency, and linear response. 
However, using conventional CCDs for high frame-rate imaging is only feasible for observing bright targets, due to the noise added in the readout process.
This has changed with the invention of the Electron Multiplying CCD, or EMCCD. 
In an EMCCD the signal from each pixel is cascade amplified before it is read out. This renders the readout noise negligible compared to the signal, and faint sources are thus also observable with the 
high frame-rate technique. 
Performing Lucky Imaging using EMCCDs is described in numerous articles \citep[e.g.][]{Mackay2004,Law2006A}, and by combining Lucky Imaging with an AO system, \citet{Law2009} were able to do diffraction limited imaging at the 5m Palomar telescope, thus reaching a spatial resolution of 35\,mas FWHM at 700\,nm.

The use of EMCCD data to do high precision photometry is an area that is just starting to be explored.
We have developed a method were the best percentage exposures are combined into a high-resolution reference frame. 
Using this reference frame in a difference imaging analysis (DIA) of other observations, including those with suboptimal resolution, makes it possible to take advantage of the best possible spatial resolution at the same time as keeping the total number of photons, in order to give best possible S/N, leading to substantial improvements in analyses of dense stellar regions.
EMCCD cameras available as off-the-shelf products provide an excellent opportunity to improve the spatial resolution, and thus perform precise time-series photometry of crowded fields on smaller telescopes.

Using EMCCD data for exoplanet detection is one of the cornerstones of the Stellar Observations Network Group (SONG) project \citep{Grundahl2006, Grundahl2009, Grundahl2014}, which will be part of the MiNDSTEp consortium operation \citep{Dominik2010} to detect exoplanets using the gravitational microlensing method. 
Microlensing which is currently the only method to improve the population statistics of cool low-mass exoplanets.
The SONG project aims at establishing a network of 1m robotic telescopes around the globe, and for these telescopes a two-colour EMCCD instrument (\tcI) has been developed.

The purpose of the \tci is to enable simultaneous red and visual band observations, thus providing instantaneous colour information and
maximising the detected wavelength range without losing spatial information.
The instrument is also equipped with a focus system that can automatically update the focus of the telescope.

To test the capabilities of the \tci before the commissioning of the SONG telescope prototype, a single EMCCD camera, and then later a full copy of the \tcI, was installed at the Danish 1.54m Telescope (DK154), at La Silla, Chile.

To run the instrument, a software system has been developed and regular EMCCD operations have been running for several years at the DK154 using this software. To the best of our knowledge, the \tci is thus the first routinely operated multi-colour instrument providing LI photometry.
This has lead to a number of interesting results, including detection of variable stars in dense globular clusters \citep{Skottfelt2013, Skottfelt2014}, and the discovery of two rings around the small asteroid-like object (10199) Chariklo \citep{BragaRibas2014}, as well as several new exoplanets already published \citep[e.g.][]{2014ApJ...782...48T} or in the process of analysis.

In Sect. \ref{sec:EMCCD} a short introduction to EMCCDs and high frame-rate imaging is given, and the EMCCD camera used for the \tci is presented. 
Sect. \ref{sec:Telescopes} gives a short introduction to the DK154 and SONG telescopes, where the TCI are used, and Sect. \ref{sec:Design} describes the design and implementation of the instrument.
The software system developed for controlling the instrument and reducing the data is described in Sect. \ref{sec:software}.
%, and Sect. \ref{sec:DataFormat} details the EMCCD data formats, as produced by the \tcI.
In Sect. \ref{sec:Performance} an analysis of the system performance of the \tci at the DK154 is given, and Sect. \ref{sec:Science} describes some of the scientific results that the \tci has achieved.

\section{EMCCDs and high frame-rate imaging} \label{sec:EMCCD}
EMCCDs, and the mathematical theory behind them, are described in great detail by in previous works \citep[e.g.][]{Mackay2001,Basden2004,Harpsoe2011,Harpsoe2012}, and only a short introduction will be given here. 

When a conventional Charged-Coupled Device (CCD) is read out, each row in the imaging area is shifted in turn to the serial, or readout, register and the pixels in each row are read-out successively through an output amplifier, where the electrons are converted into ADUs (Analog to Digital Unit). 
The lowest noise output amplifiers can determine the charge with a precision down to a few photons, but if one wants to speed up the readout frequency, the noise in the readout increases to tens or hundreds of electrons.
It is therefore undesirable to do high frame-rate imaging of faint sources with a conventional CCD, as the signal would be lost in readout noise.

An EMCCD has an extended readout register, or EM register, that amplifies the signal before it is read out. This is done by shifting the charges in each step of the EM register with a significantly higher voltage than normally.
Each time an electron is shifted, there is a small probability $p_m$ that an impact ionisation will occur, resulting in an extra electron. This creates a cascade amplification over the length of the EM register.

The amplification of a single electron over $m$ steps in the multiplication register, is given by the multiplicative gain 
\begin{equation}
\gamma = \left( 1+p_m \right)^m \ . \label{eq:gamma}
\end{equation}
Thus, with a multiplication register of $m=600$ steps, and $p_m = 0.01$, the gain is $\gamma = 1.01^{600} \approx 400 \, \elec/\text{photon} $.

Based on Eq. \ref{eq:gamma}, the probability density function (PDF) for a certain number of output electrons $x$ given a number of input electrons $n$, can be estimated as
\begin{equation}
P(X=x \,|\, n) = x^{n-1}\frac{e^{-x/\gamma}}{\gamma^n \Gamma(n)} \ , 
\label{eq:erlang}
\end{equation}
where $\Gamma$ is the gamma function.
This distribution is known as the Erlang distribution and it is a special case of the gamma distribution. 

   \begin{figure}
   \centering
   \includegraphics[width=\linewidth]{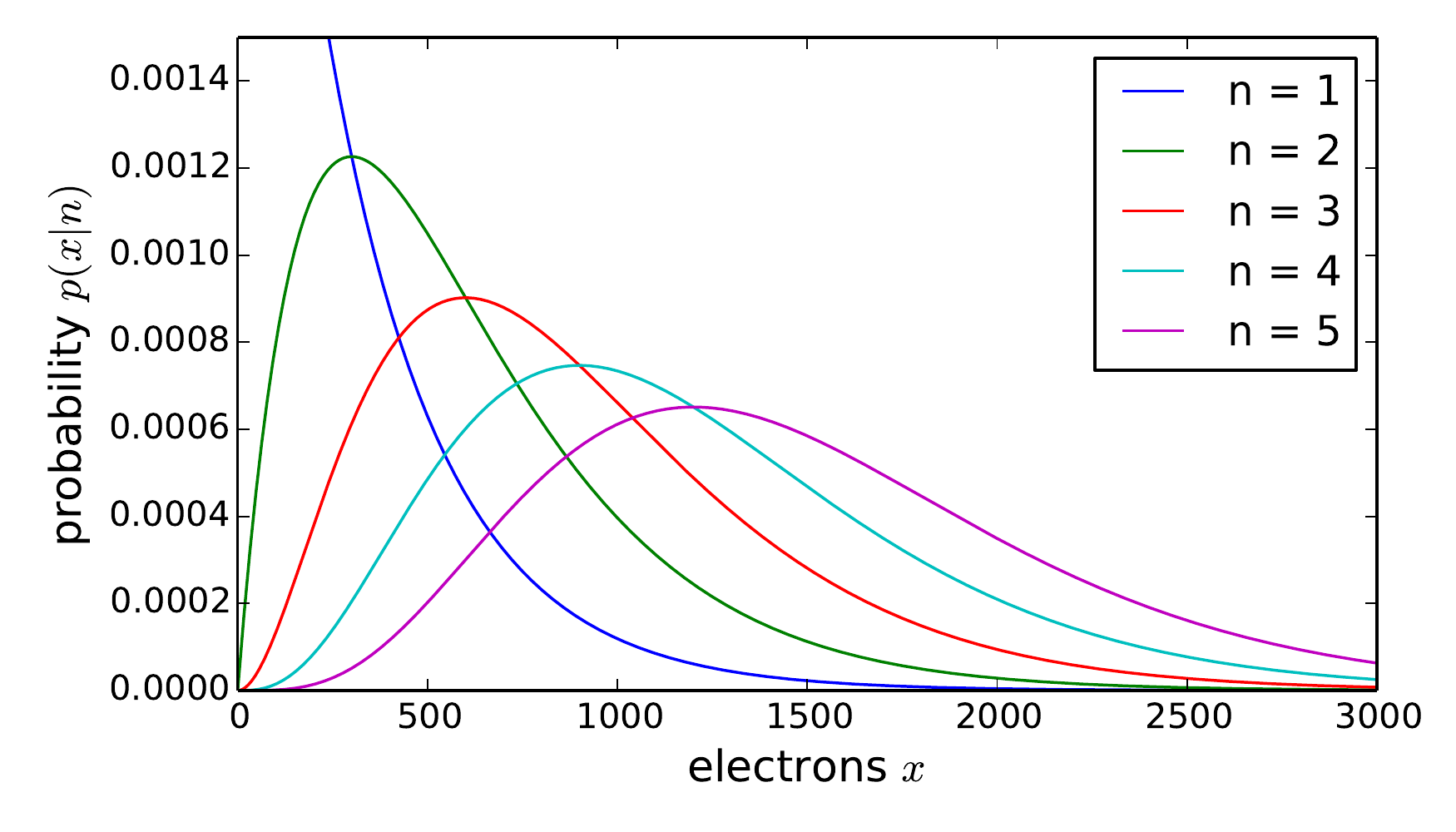}
      \caption{Plot of the Erlang distribution as defined in Eqn. \ref{eq:erlang} and for $\gamma=300\,\elec/\text{photon}$. As $\gamma$ is a scale parameter, it will only scale the numbers on the x-axis. It will not change the functional form \citep[Adopted from][figure 2]{Harpsoe2012}.}
         \label{fig:erlang}
   \end{figure}
   
A plot of the Erlang distribution for different values of $n$ is shown in Fig. \ref{fig:erlang}, illustrating that there are large overlaps between the possible outcomes. 
In conventional CCDs the photon noise, or shot-noise, is given by the Poisson distribution, which means that the signal-to-noise ratio can be determined as $\text{S/N} = S/\sqrt{S}$. 
The stochastic nature of the cascade amplification contributes an extra factor of 2 to the variance, usually known as the excess noise factor. The S/N for the EMCCD therefore becomes 
\begin{equation}
\text{S/N} = \frac{S}{\sqrt{2S}} \ .
\end{equation}

Another drawback of the EMCCD is the signal coming from spurious charges. 
Spurious charges are electrons that arise in the CCD during the readout, as an effect of the vertical or horizontal shift operations.
These events are usually rare (they only happen in a few per cent of the pixels) and release a single electron in each event. In conventional CCDs they are undistinguishable from the readout noise. 
However, for an EMCCD where all electrons are cascade-amplified, spurious charges give rise to a detectable signal.
Unless the EMCCD is used for photon-counting purposes, the signal from the spurious charges is assumed to be negligible.

With the signal amplified in the EM register, for instance by a value of $\gamma=300\,\elec/\text{photon}$, the readout noise becomes negligible even at high readout speeds. 
EMCCDs thus make it feasible to do high frame-rate imaging, which has a number of advantages that we will discuss in the following.

%The main advantage is that by 
By shifting and adding the individual exposures, it is possible to compensate the blurring effects of atmospheric disturbances and achieve a better spatial resolution. 
This method is known as speckle imaging, or just shift-and-add.
\citet{Fried1978} showed that the probability of obtaining a good short exposure (i.e. an exposure where the wavefront area over the aperture is less than 1 rad$^2$), is
\begin{equation}
P = 5.6\exp\left(-0.1557 (D/r_0)\right)\quad (\text{for}\, D/r_0 \geq 3.5),
\end{equation}
where
% $D$ is the telescope diameter and 
$r_0$ is the Fried parameter.
The Fried parameter is a measure of the optical aberrations caused by atmospheric disturbances and is defined as the diameter of a circular area over which the rms of the wavefront aberrations is 1 rad.
The Lucky Imaging technique makes use of this theory, by only stacking the few per cent best frames to obtain very high spatial resolution. Shifting-and-adding some percentage of the best frames is referred to as \emph{percentage selection}.

\begin{figure}
   \centering
   \includegraphics[width=\linewidth]{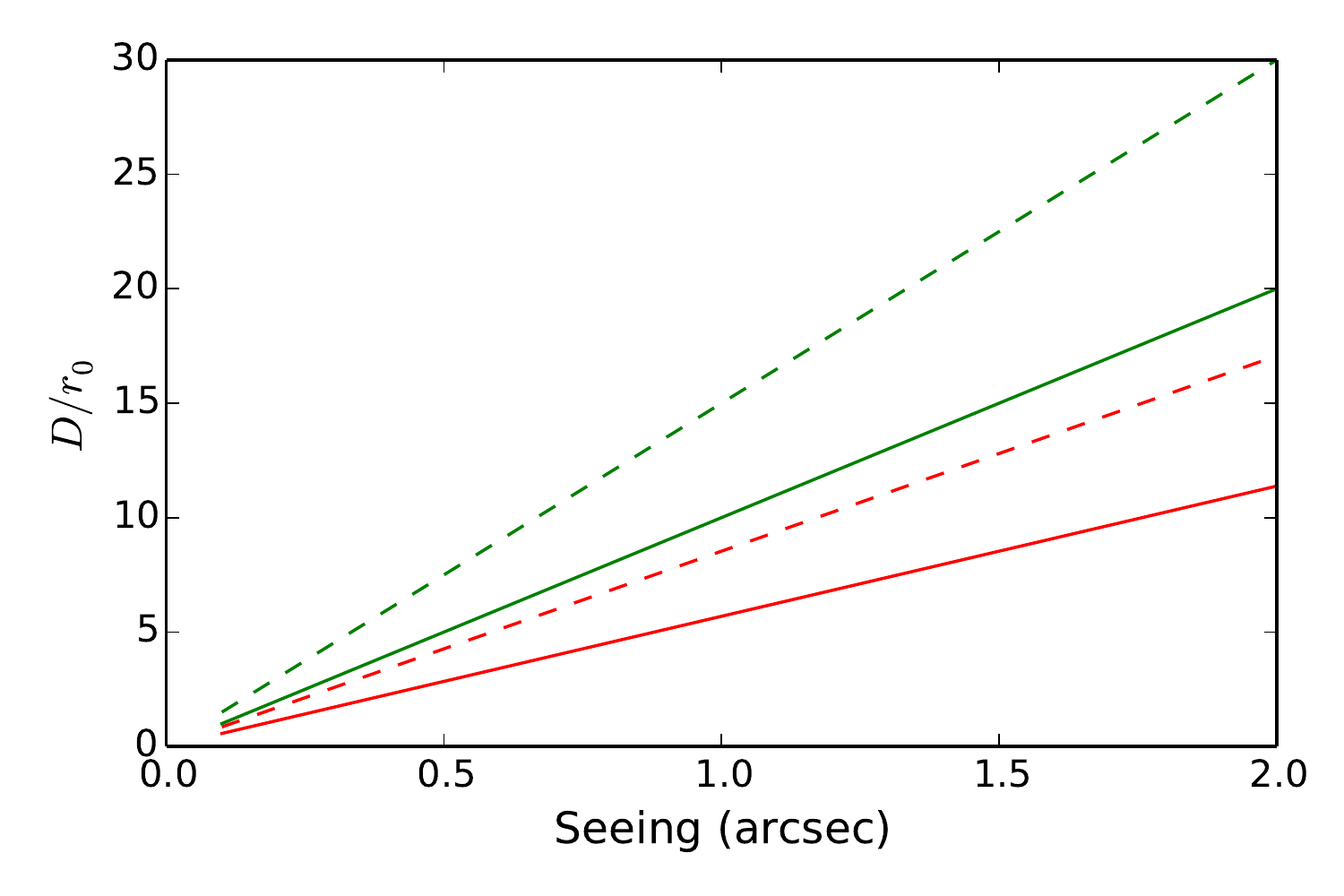}
      \caption{Plot of $D/r_0$ as function of seeing is plotted for 500 nm (green) and 800 nm (red) for a 1 m (solid line) and 1.5 m (dashed line) telescope.}
         \label{fig:D_r0}
   \end{figure}

The improvement of the Strehl ratio using high frame-rate imaging was examined by \citet[hereafter S09]{Smith2009}. Strehl ratio is a measure of image quality defined as the ratio of the peak intensity in an observed image, to the maximum attainable intensity in a diffraction limited system. 
Using standard $V$ and $I$-band observations of bright stars, S09 examined $D/r_0$ ratios between 3 and 30. 
%where $D$ is the telescope diameter and $r_0$ is the Fried parameter \citep{Roddier1981}.
%The Fried parameter is a measure for the optical aberrations caused by atmospheric disturbances, and is defined as the diameter of a circular area over which the rms wavefront aberrations is 1 radian.
For comparison, a plot of $D/r_0$ versus seeing for a 1 and 1.5 m telescope, is shown in Fig. \ref{fig:D_r0}.

At short exposure times, below 10\,ms, S09 found that the Strehl ratio of the shift-and-added frames is between 2 and 3 times higher than stacked frames with no shifting. 
For a 1\% selection the improvement is between 4 and 6, with a peak at $D/r_0 \sim 7$. 
The effect of the single frame exposure times, $t$, was also examined, though only for a 1\% selection. 
Here the shortest exposure time, $t = 1\,\text{ms}$, gave the best results with Strehl ratios of up to six times higher.
At longer exposure times the improvement decreased, but even at $t = 640\,\text{ms}$, the Strehl ratio was improved by a factor of two.

The very short exposure time that is used when doing high frame-rate imaging makes it possible to observe even quite bright objects without saturating the CCD. 
The required S/N for the fainter objects can then be reached by combining the required number of exposures at a later stage, thus achieving a very high dynamical range. 
This makes it possible to perform high precision photometry for both bright and faint stars in the same field, something that can be hard to obtain when doing conventional CCD imaging.

\subsection{EMCCD camera} \label{sec:cam}
The Andor iXon+ 897 EMCCD was chosen as the detector for both the red- and the visual-band of the Two-Colour Instrument (\tcI). 
The camera has an active imaging area of $512\times512$ pixels, and with the chosen pixel scale of $0\farcs09/\text{pix}$, this gives a field-of-view (FoV) of about $45\arcsec \times 45\arcsec$. 
The EM readout amplifier has a maximum pixel readout rate of 10 MHz, and is thus able to readout $\sim35$ full frames per second. 

The typical readout noise at maximum readout rates is $\sim 50\,\elec$, but by using the EM gain, which can be as high as $\gamma=1000\,\elec/\text{photon}$, the readout noise becomes negligible compared to the signal even at a single photon level, and the camera can thus be said to be photon counting. 
However, to protect the camera, Andor suggest that the EM gain is not set higher than $\gamma=300\,\elec/\text{photon}$ in normal operations, and that EM gain values above this are only used on special occasions. 

The camera has a thermoelectric (Peltier) cooling system that can cool the camera down to $-85\degr$C. At this temperature the typical dark current in the camera is 0.001 $\elec$/pix/sec. The typical spurious charge rate is 0.005 $\elec$/pix/sec.

The camera is also equipped with a conventional readout amplifier with pixel readout rates of 3 or 1\,MHz, and with a typical readout noise of 6 electrons for the latter.

The wavelength coverage goes from 300\,nm to 1100\,nm. The quantum efficiency (QE) for the camera is discussed in more detail in Sect. \ref{sec:QE}.
%The quantum efficiency (QE) curve for the camera is plotted in Fig. \ref{fig:QE}. 
%The spectral responce of the whole system is discussed in more detail in Sec. \ref{sec:QE}.

\section{Telescopes} \label{sec:Telescopes}
The \tci was originally designed for the SONG telescope prototype, but in order to test the capabilities of the \tcI,
% until the SONG prototype came online, 
a copy was made for the DK154.
A short introduction to the telescopes is given here.

\subsection{Danish 1.54m Telescope}
The DK154, built by Grubb-Parsons, was commisioned in 1979. 
The telescope is situated at the ESO La Silla Observatory, Chile (\Geo{70}{44}{662}W \Geo{29}{15}{14}{235}S), at an altitude of 2340 metres.

Since its commissioning, the telescope has been refurbished a number of times. 
In 2012, a complete overhaul of the electrical and mechanical systems was done by the Czech company Projectsoft\footnote{www.projectsoft.cz}. 
A new telescope control system (TCS), also developed by Projectsoft, and other subsystems were also installed. 
It is thus now possible to control the telescope remotely, such that remote observations without on-site staff can be done. 

The telescope has an off-axis equatorial mount and the optics are of a Ritchey-Chr\'etien design. 
The main instrument is the Danish Faint Object Spectrograph and Camera (DFOSC), which was installed in 1996 \citep{Andersen1995}, though since 2004 scaled down to exclude the spectroscopic capability. 
The DFOSC uses a 2K$\times$2K thinned Loral CCD chip with a pixel scale of $0\farcs4/\text{pix}$, exhibiting a FoV of $13\farcm7 \times 13\farcm7$.
%The DFOSC is mounted on an adaptor, that houses an auto-guiding system and two filter wheels. 

%In 2009 a single Andor iXon+ 897 EMCCD camera was installed at an auxiliary port on the telescope, to be used as a testbed for the two-colour instrument.
%The camera was therefore equipped with a special long-pass filter with a cut-on wavelength of 650 nm (Thorlabs FEL0650) to mimic the transmission of the red-vis dichroic mirror in the \tci. 

\subsection{SONG}
The Stellar Observations Network Group (SONG) project aims at constructing a network of fully robotic 1m telescopes for doing time-domain astronomy, in particular asteroseismic observations and exoplanet studies \citep{Grundahl2006,Grundahl2009, Grundahl2014}.
To ensure the continuous coverage of both northern and southern targets, eight telescope nodes are required, and these should be placed on existing observatory sites around the globe. 

The prototype for the SONG telescope is delivered by the German company Astelco\footnote{http://www.astelco.com/}, while its two instruments are assembled at the workshops at Aarhus University (the spectrograph) and at University of Copenhagen (the \tcI). The telescope is carried by an Alt-Az mount driven by magnetic torque motors ensuring fast pointing (up to $20\deg$ per second) with high precision (within $3\arcsec$ of the pointing model). 

A detailed description of the SONG optical system, including schematics, is given in \citet{Grundahl2009}, so only a short introduction will be given here. 
The 1m primary mirror is only 5 cm thick and employs an active correction system based on Shack-Hartmann measurements of bright stars. With this system the telescope will have nearly diffraction limited optics. 

%Optical derotator and atmospheric dispersion corrector (ADC) and path selector.

A high resolution spectrograph is installed at the Coud\'{e} focus. This will be used to do high precision radial-velocity measurements. 

The \tci is installed at the Nasmyth focus and is part of the Nasmyth unit that also contains a de-rotator and an atmospheric dispersion corrector (ADC).

To take advantage of the diffraction limited optics, and the high frame-rate capabilities of the EMCCD cameras, a sampling of $0\farcs09/\text{pix}$ was chosen. This will ensure a sampling better than 2 pixels of the $I$-band diffraction limit. 
With a seeing of $1\arcsec$, the $D/r_0$ ratio for the SONG telescope is between 5 and 6 at a wavelength of 800 nm. As described in Sect. \ref{sec:EMCCD}, this is near the optimal regime for the Lucky Imaging method.

The SONG telescope prototype has been installed at the Observatory del Teide on Tenerife, Spain, and a second SONG node has been built in China.

\section{Two-Colour Instrument} \label{sec:Design}
The motivation behind the Two-Colour Instrument was that it would provide instantaneous colour information about the observed objects and maximise the detected wavelengths without losing spatial information. 

%The idea behind a two-colour instrument was to maximise the detected wavelengths without loosing spatial information. It would also be able to provide instant colour information about the detected objects. 

In 2009, a single Andor EMCCD camera was installed at an auxiliary port on the DK154, to be used for testing the capabilities of the \tcI, in anticipation of the SONG commissioning.
The camera was therefore equipped with a special long-pass filter with a cut-on wavelength of 650 nm (Thorlabs FEL0650), to mimic the transmission of the red/vis dichroic mirror in the \tcI, and had a sampling of $0\farcs09/\text{pix}$.

After a few years of successful testing it was decided in late 2012, to build and implement a full version of the \tci on the DK154. 

\subsection{Optical design}
The \tci basically consists of two dichroic mirrors, two EMCCD cameras, and a focus system.
The first (red/vis) dichroic mirror transmits the red band to one EMCCD camera (the 'red' camera), and reflects the visual-blue band towards the second dichroic mirror. 
Here the visual band is transmitted to the second EMCCD camera (the 'visual' camera), and the blue band is reflected towards the focus system. The focus system is described in Sect. \ref{sec:focus}.

\begin{figure}
   \centering
   \includegraphics[width=\linewidth]{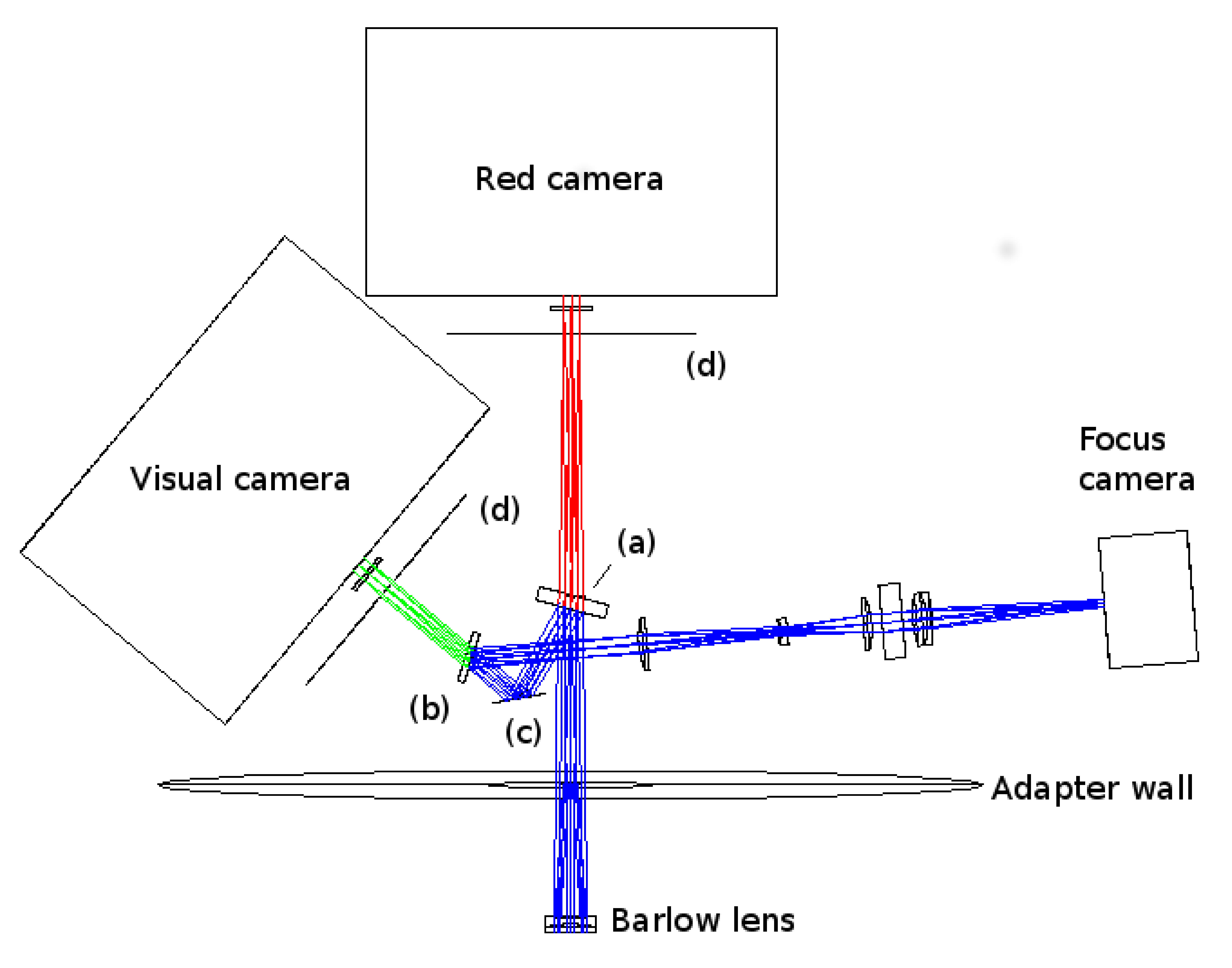}
      \caption{Optical design for the \tci at DK154. Here (a) is the red/vis dichoic, (b) is the vis/blue dichroic, (c) is the folding mirror, and (d) marks the filter wheels.}
         \label{fig:optical_design}
   \end{figure}

In order to achieve the same sampling on the DK154 as on the SONG telescope, a Barlow lens is added to the system before the red/vis dichroic mirror. 
It was also necessary to add an extra folding mirror on the DK154 \tci after the red/vis dichroic mirror and to rearrange the cameras, in order to create a system that could be mounted at the auxiliary port on the DK154.
The optical design for the \tci on the DK154 is shown in Fig. \ref{fig:optical_design}.

The SONG \tci has a filter wheel in front of each EMCCD camera. 
In the DK154 \tci there are no filter wheels installed, but room has been allocated for them, so that they can be installed at a later stage. 

\subsection{Instrument efficiency} \label{sec:QE}
\begin{figure*}
   \centering
   \includegraphics[width=\linewidth]{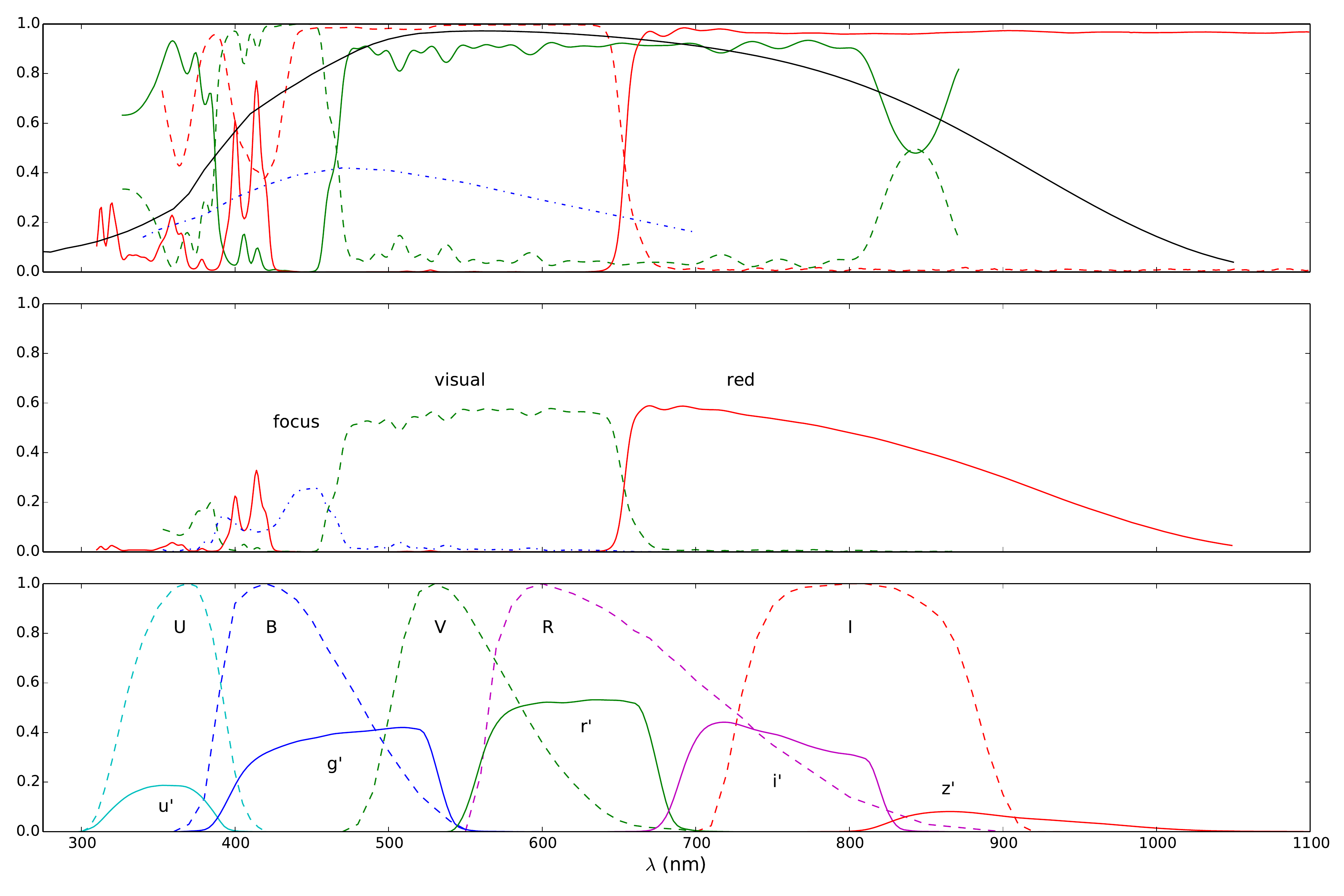}
      \caption{Top panel shows the transmission (solid) and reflection (dashed) of the red/vis (red) and vis/blue (green) dichroics, respectively, the QE of the EMCCD camera (solid black), and the QE of the focus camera (dot-dashed blue). 
      Middle panel shows the instrument efficiency of the red (solid red line), visual (dashed green line), and focus channel (blue dot-dashed line). 
      In the bottom panel the normalised passbands of the Johnson-Cousins $UBVRI$ photometric system (dashed lines) are plotted along with the system sensitivity of the SDSS $u'g'r'i'z'$ camera (solid lines). The latter is based on a sensitivity determination made by Jim Gunn in June 2001, which is available as a FITS file at \texttt{http://www.sdss3.org/instruments/camera.php}. } 
         \label{fig:QE}
   \end{figure*}

The efficiency of the \tci has been calculated based on the supplied transmission/reflection curves of the dichroics and assuming that the telescope has a transmission efficiency of $65\%$. 
For the \tci on the DK154, a small loss in the Barlow lens ($\sim 3\%$) would have to be taken into account, but this has not been included here. 
%The efficiency of the \tci is mainly defined by the cross-over wavelengths $\lambda_c$ of the dichroics and the quantum efficiency (QE) of the cameras.
%Losses in the telescope optics will also affect the system efficiency, but this has not been included here. 
The plot in the top of Fig. \ref{fig:QE} shows the reflection and transmission efficiency
% ($\eta_r$, and $\eta_t$, respectively) 
of the dichroics, the QE of the Andor EMCCD camera, and the QE for the Sony ICX285 CCD chip used by the focus camera (see Sect. \ref{sec:focus}).

In the middle panel of Fig. \ref{fig:QE} the relevant efficiencies are multiplied with the QE of the cameras and the transmission efficiency of the telescope to give the instrument efficiency for the three channels.
The lower panel of the figure shows the normalised passbands of the Johnson-Cousins $UBVRI$ photometric system \citep{Bessell1990} and the sensitivity of the Sloan Digital Sky Survey (SDSS) camera through the $u'g'r'i'z'$ filter system \citep{Gunn1998} for comparison purposes.

The passbands of the TCI are defined by the dichroics and are significantly wider than the standard photometric filter systems, with effective bandwidths of $\sim 250\,\text{nm}$ and $\sim 185\,\text{nm}$ in the red and visible channel, respectively.
   
The red/vis dichroic has a cross-over wavelength $\lambda_c^r \simeq 655$\,nm and the red channel therefore corresponds roughly to a combination of the SDSS $i'+z'$ filters. 
There is a leak in the dichroic between 350-430\,nm, which will influence the photometric output. The effect of this can be eliminated by using a long-pass filter in front of the red camera.

With a $\lambda_c^v \simeq 466$\,nm for the vis/blue dichroic, the visual channel does not match any of the standard photometric systems. 
The cut-on wavelength is right in the middle of the SDSS $g'$ filter, and the cut-off wavelength $\lambda_c^r \simeq 655$\,nm is about 40\,nm from the red end of the SDSS $r'$ filter at 695\,nm.
There is also a contribution to the visual channel for wavelengths shorter than 390 nm. Again this can be eliminated by using a (long-pass) filter in front of the visual camera.

Only a small part of the light, $\sim 390-460$\,nm, is reflected towards the focus system by the vis/blue dichroic. The loss in the focus optics is not accounted for, but the efficiency below $\lambda = 400$\,nm is usually poor, and combined with the leak in the red/vis dichroic, this means that the focus system will mainly use wavelengths between 420 and 460\,nm.  %This matches the Johnson $B$ filter quite well.

The displayed spectral responses for the red and visual channels can of course be changed if filters are added to the system. This could, for instance, include filters for standard photometric systems in order to obtain a colour output that is more comparable to other studies.

%It has been discussed to change $\lambda_c$ for the two dichroics, such that the channels would better match, for instance, the SDSS passbands.

\subsection{Focus system} \label{sec:focus}
For the EMCCD cameras to reach their full potential it is important that the telescope is in the correct focus at all times.
The \tci therefore has a focus system that uses the light reflected by the vis/blue dichroic. 
The light is collimated and sent through a four-quadrant wedge 'focus pyramid'. 
When the light from the focus pyramid is focussed onto the focus camera by the camera lens, each quadrant will be slightly displaced, and the distance between the spots depends linearly on the focus offset. 
By using a script that can analyse the focus data and calculate the current focus offset, it is possible to do an automatic focus correction of the telescope.

A Prosilica GC1350 CCD camera from Allied Vision Technologies, using the Sony ICX205 CCD chip, was initially chosen as a focus camera.
%Unfortunately, only a small wavelength range, $\sim 420-450$ nm, is available for the focus system, and combined with the relatively high readout noise this limits the strength of the detected signal. % (see Sect. \ref{sec:QE}).
However, as only a small wavelength range ($\sim 420-460\,\text{nm}$) is available for the focus system, and since the Prosilica camera has no cooling system and a rather high readout noise level, it was not possible to detect a reasonable signal even for relatively bright stars.

After analysing the problem, it was found that the Atik 314L+ camera solves the problem. This camera uses the Sony ICX285 CCD chip, and has a cooling system and a lower readout noise level of around $5\,\elec$. 
Calculations show that with this CCD camera using 2x2 pixel binning, an integration time of 10\,s, and a seeing of $0\farcs5$, it should be possible to focus the telescope if there is a star brighter than about $m_B = 17$\,mag in the field. 
Based on \citet{Allen1973}, we find that with this limit, the sky coverage is over $70\%$ on average and over $95\%$ near the Galactic plane. 
At a seeing of $1\arcsec$ the magnitude limit is $m_B \simeq 16.5$\,mag, and the sky coverage thus drops to $60\%$ on average and $90\%$ near the Galactic plane.
It should be noted that the data from \citet{Allen1973} used for these calculations is criticised in \citet{Bahcall1980} for up to a 40\% too high density.

\subsection{Technical Implementation}
\begin{figure}
   \centering
   \includegraphics[width=\linewidth]{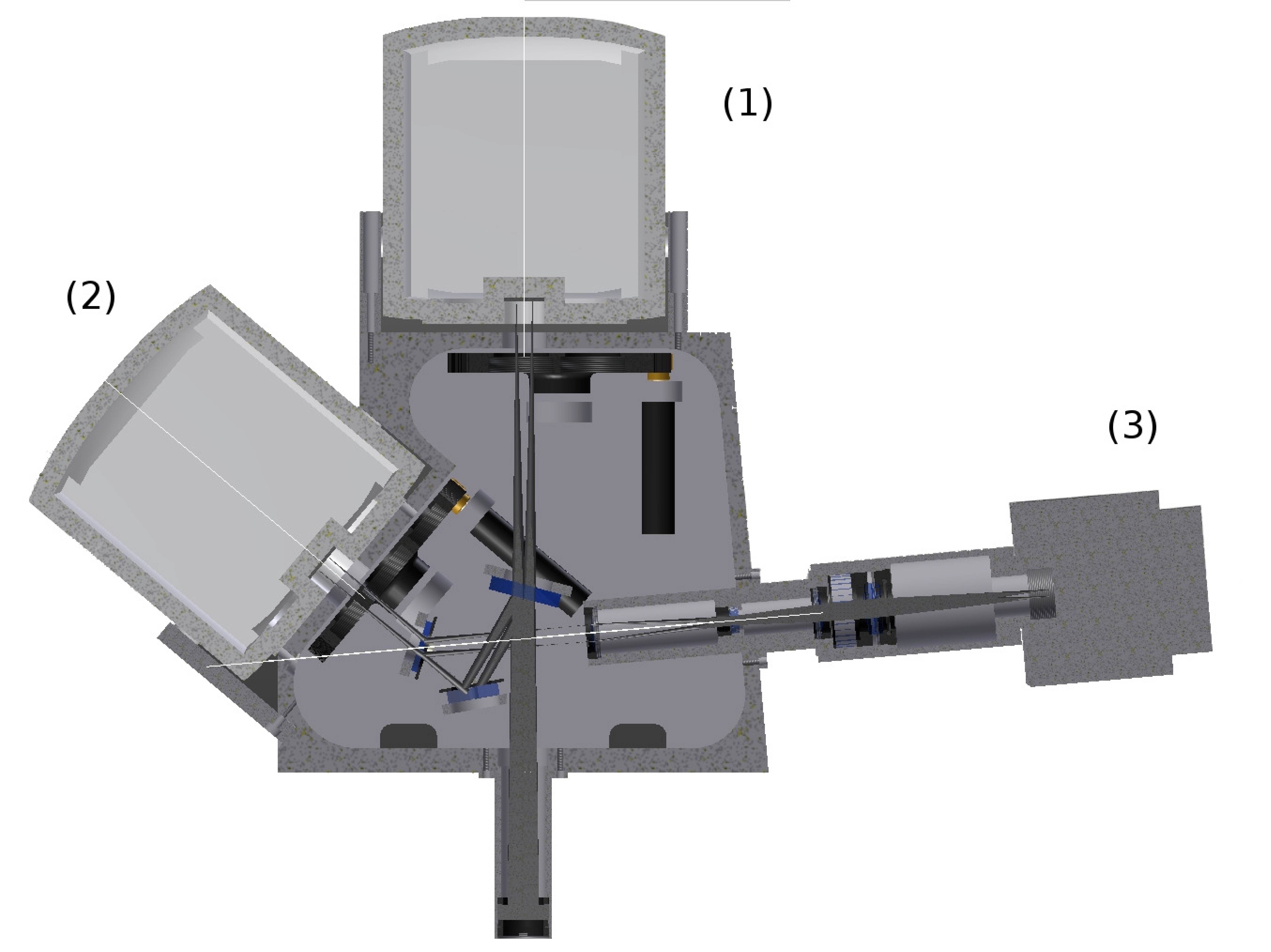}
      \caption{Technical drawing of the \tci at DK154. Attached to the instrument housing are (1) the red, and (2) the visual EMCCD cameras, and (3) the focus camera. Filter wheels are included in the drawing, although these are currently not installed. }
         \label{fig:technical}
 \end{figure}

A technical drawing of the instrument is shown in Fig. \ref{fig:technical}. 
This figure also shows the positions of the filter wheels inside the instrument housing. 

Andor has developed a special PCI controller card to control the EMCCD cameras. Each camera thus needs to be connected with the PCI controller-card using a two metre long signal cable. 

On the SONG telescope, the Andor PCI cards are installed in two lightweight computers attached directly on the telescope mount, which transfer the data from the cameras to a server using a gigabit ethernet connection.

\begin{figure}
   \centering
   \includegraphics[width=0.9\linewidth]{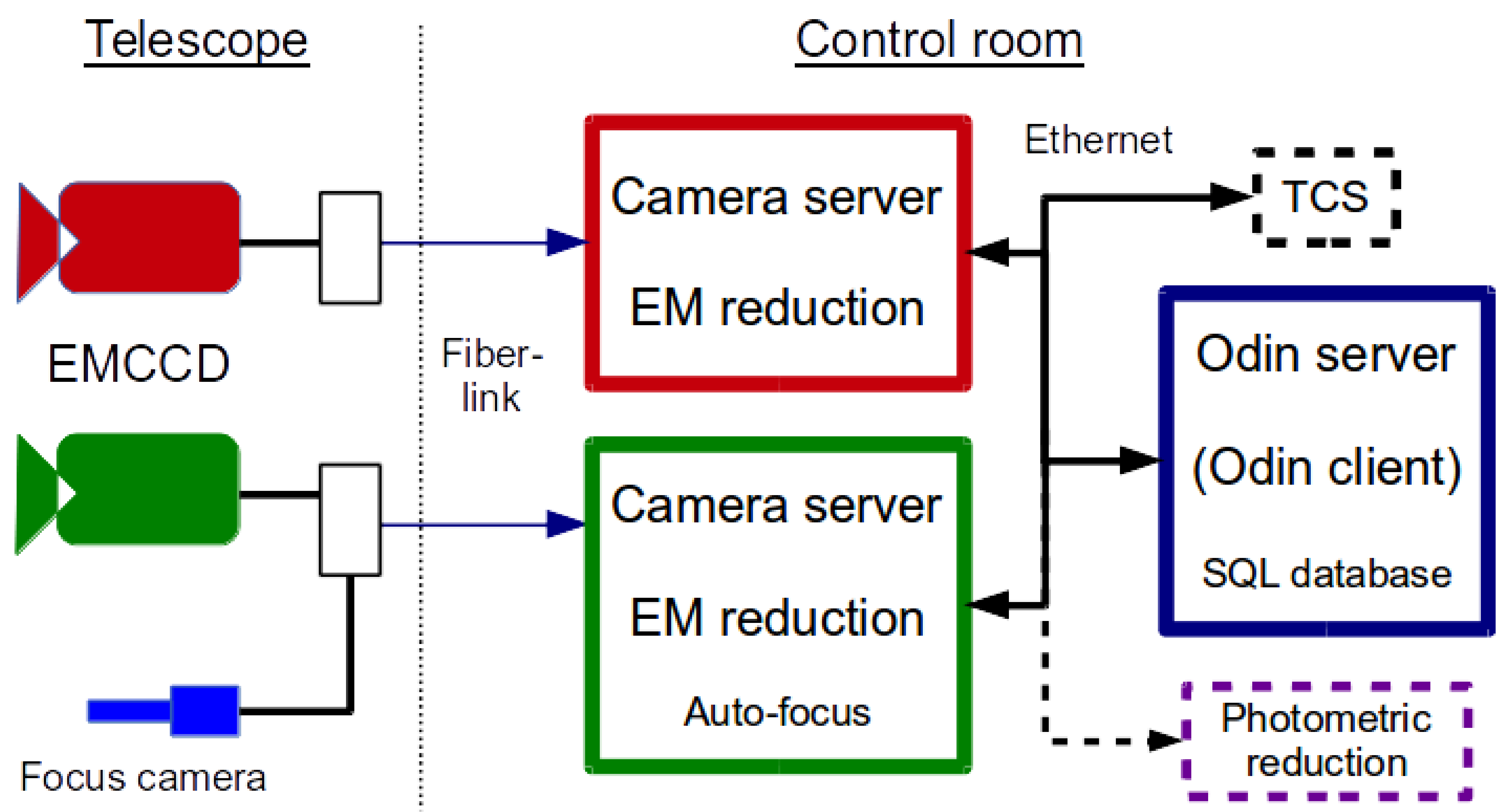}
      \caption{Overview of the \tci and the Odin system at DK154. The EMCCD cameras, and the focus camera, are attached to two cabinets, which each are connected to a computer, running the camera server and EM reduction software. The camera servers are controlled by the Odin server running on a third computer. The Odin server also handles the communication with the TCS. The Odin server can be controlled by the Odin client, which can be operated either locally at the telescope or remotely. After the data are reduced by the EM reduction software, they are ready for a photometric reduction.}
         \label{fig:system}
 \end{figure}

On the DK154 we have chosen to use the Adnaco\footnote{http://www.adnaco.com} S1A PCI/PCIe fiber-optics extension system. The Adnaco system consists of 
\begin{enumerate}[(a)]
\item a small motherboard, with two PCIe and two PCI slots, in one of which the Andor controller-card is attached,
\item a PCIe card, that is installed in a computer in the control room, and 
\item a 50 metre fibre cable connecting (a) and (b).
\end{enumerate}
The control room computers are now able to use the PCI/PCIe slots on (a), and thus the Andor controller card, as if they were installed directly in the computer. 
Using this system ensures a fast and very stable connection to the EMCCD camera and physical access to the computer running the camera even when the telescope is being operated.
It also eliminates any mechanical wear that would otherwise have been caused on hard disks (HDDs) and other moving parts attached to a moving telescope. 
The focus camera is attached to a USB PCIe card that is connected to a PCIe slot on (a). 
This makes it possible to run the focusing routine described in Sect. \ref{sec:focus} on the control room computer. 
See also the schematic in Fig. \ref{fig:system}.

\begin{figure}
   \centering
   \includegraphics[width=\linewidth]{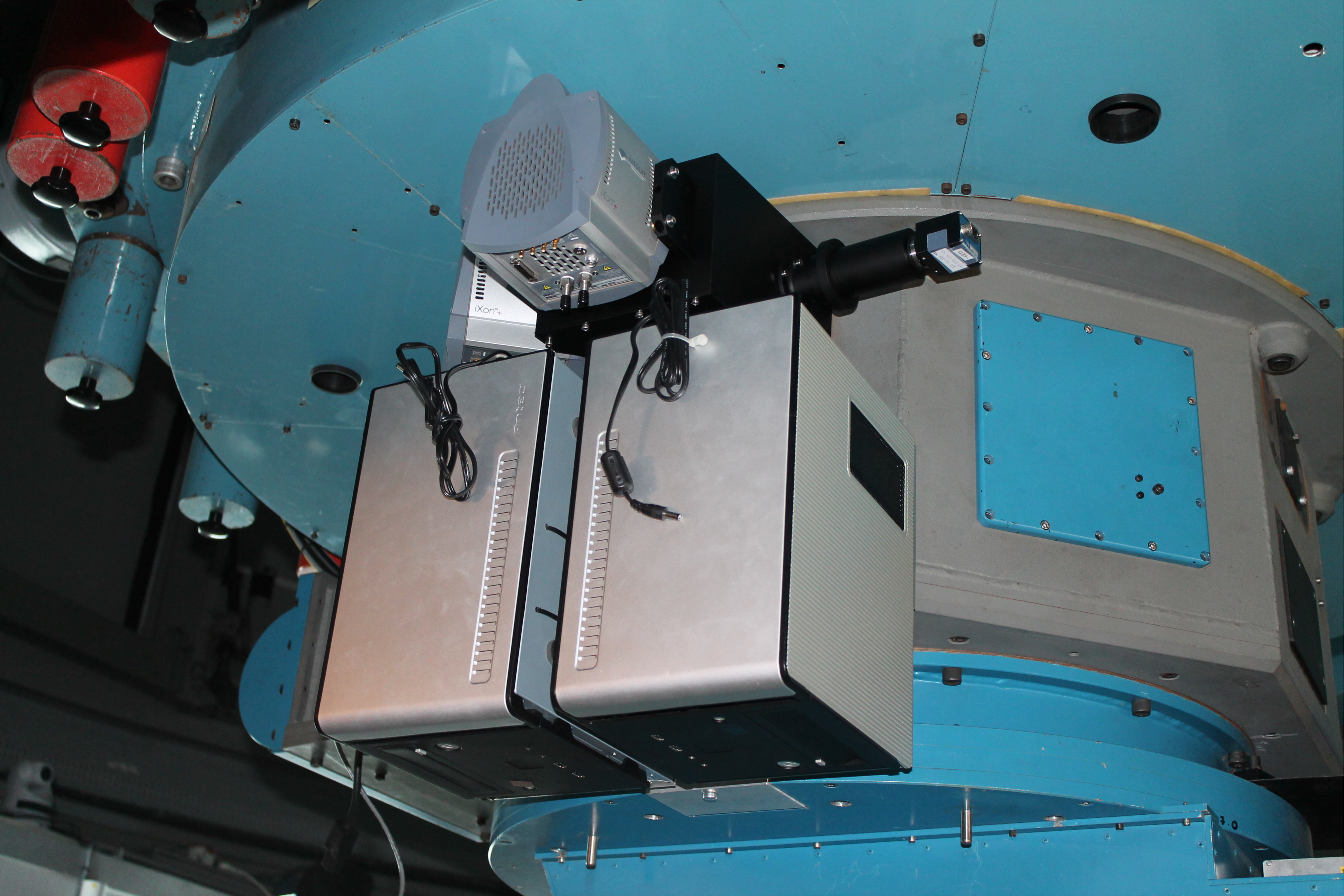}
      \caption{Picture of the \tci installed on DK154. Beneath the instrument housing and the cameras are the two cabinets that contain the Adnaco fiber-link motherboards, and the Andor PCI cards.}
         \label{fig:installed}
 \end{figure}

Fig. \ref{fig:installed} is a picture of the \tci installed on DK154. Here the two cabinets containing the two Adnaco and Andor cards, one for each EMCCD camera, can be seen attached to the telescope beneath the instrument housing and cameras.

Each of the cameras are run by a HP Compaq Elite 8300 PC, employing a quad-core processor and 24\,GB RAM. The computers are running the Ubuntu 12.04 LTS distribution of Linux. 
The computers are responsible for acquiring, saving, and processing the high frame-rate data. At a 10\,Hz frame-rate, this is about 300\,MB/min, and thus up to 200\,GB/night/camera. 
Besides a 500\,GB HDD for the control system and other software, the computers are therefore equipped with both a 1\,TB and a 4\,TB HDD.

The \tci was installed at the DK154 in the spring of 2013. There were some hardware problems delaying the operations of the visual camera. These problems were fixed in the summer of 2014, and both EMCCD cameras are now working. 
%The focus camera was updated in Autumn 2014 and initial tests have shown that this can be made operational, too.

\section{Software} \label{sec:software}
A new software system, called \textit{Odin}\footnote{\textit{Odin} (or W\=oden or Wotan) is the Alfather of the gods from the Norse mythology. He is one-eyed and uses his two ravens, Huginn and Muninn (''Thought'' and ''Memory''), to keep an eye on the world.}, has been developed for the \tcI.

The Odin system makes use of a server-client structure in order to run the two cameras simultaneously, and to enable it to run on different hardware set-ups. 
The system has built-in communication with the TCS, enabling it to do automated observations, focus adjustment etc. 
It has an automatic pointing correction routine, that ensures a pointing precision of better than $0\farcs5$ for previously observed targets. The latter is especially important given the relatively small FoV of $45\arcsec \times 45\arcsec$.

The software system for an instrument like the \tci will always evolve, as more and/or different functionality is required. 
The Odin system has evolved a lot, and will most likely evolve in the future. 
However, using some version of this software has made it possible to carry out regular EMCCD observations at the DK154 for several years. 

\subsection{Overview}
The  Odin system consists of a number of modules that are running as daemon (or background) processes:
\begin{itemize}
\item The \textbf{camera servers} handle the communication with the cameras and save the raw data. 
\item The \textbf{Odin server} handles the communication with the camera servers and the TCS.
\item The \textbf{EM reduction software} calibrates the raw data, and saves it as FITS files. 
\item An \textbf{SQL database} is used for storing, and sharing information about the telescope and camera status between different parts of the 
system. 
\item The \textbf{autofocus script} is described in Sect. \ref{sec:focus}.
\end{itemize} 

For the DK154 a graphical user interface (GUI), called the Odin client, has been implemented for controlling the Odin server. The Odin client can be used either locally at the telescope, or from a remote location, depending on the physical location of the observer.
For the SONG telescope, which is fully robotic and therefore does not have an observer present, a virtual observer module will control the Odin server.

An overview of the \tci and the Odin software is given in Fig. \ref{fig:system} and the different parts of the system will be described in more detail below.

We have chosen to implement the system in Python, due to its widespread use in astronomy, rapid development capability, and large number of existing modules. 
To communicate between the servers we use the \texttt{Pyro} (PYthon Remote Objects) module, as this is completely written in Python, stable, and easier to use than, for instance, XML-RPC. 
Due to the complexity of the system, and since many tasks need to be performed in parallel, the \texttt{multiprocessing} and \texttt{threading} modules have been used to create parallel processes and threads, and the \texttt{Queue} module is used to interact between them.
 
An elaborate error handling system has been implemented, such that all errors will be caught and dealt with, either by the system itself or via a message to the observer through the GUI. 
This part of the software is crucial as the system, in case of an error, would otherwise be left in a state that would require a complete reboot.

To make the system as versatile as possible we have made an effort to gather the telescope and/or site specific parameters in a single configuration file, and have arranged the TCS communication functions in a single module.
This means that only very few files need to be telescope specific, making the Odin software easier to implement and maintain at multiple telescopes.

\subsection{Camera servers} \label{sec:camServer}
The main purposes of the camera servers are to control the relevant camera and to save the data that it produces.
Andor delivers a Linux driver for their cameras in the form of a shared library, which can be controlled from Python using the \texttt{pyandor} project\footnote{http://code.google.com/p/pyandor/}. 
This means that the Andor cameras can be controlled directly from Python and the data from the cameras can be read directly into \texttt{NumPy} arrays. 

The camera servers contain all the functions that are needed to change the settings of the camera, such as gain, temperature, etc., and these changes are reported to the SQL database. 
Whenever an acquisition is started (and ended), the SQL database is updated with information about the type of acquisition, the starting time, etc., so that this information is available to the GUI.
The server also has an internal lock, ensuring that an acquisition cannot be started while another is running. 

\subsubsection{Calibration data}
The camera server also includes routines for obtaining calibration data, i.e. bias and flat-field correction frames for the EM readout register.

A master bias frame $B(i,j)$ is created by reading out $k=1000$ frames
%, each containing $M$ pixel columns and $N$ pixel rows, 
using the chosen gain.
In order to remove the contribution from spurious charges, the 5\% highest values for each $(i,j)$ pixel over all $k$ frames are rejected. For the remaining pixel values the mean value over each $(i,j)$ pixel is calculated. This corresponds to the 5\% truncated mean from above. Using this method, we find that the mean variation in pixel value between the master bias frames is $\sim 0.25$ ADU, which corresponds to $\sim0.01$\,photon.

The flat-field frame $F(i,j)$ is meant to correct for gradients caused by variations in the pixel-to-pixel sensitivity of the EMCCD and attenuations in the optical path, and not effects from the EM gain, which might change from exposure to exposure. 
The flat-field frames are therefore done without any gain, i.e. $\gamma=1\,\elec/\text{photon}$. 
This means that a special master bias frame with $\gamma=1\,\elec/\text{photon}$ is needed for the flat-field frames, and a routine for this has been made. 

Both a dome flat-field and a sky flat-field routine are available for the user.
For the dome flat routine, the telescope should be pointed towards a uniformly illuminated screen attached to the dome and $k=20$ exposures are made.
The sky-flat routine needs to be run in either evening or morning twilight, with the telescope pointed at a part of the sky with as few stars as possible. 
A start exposure time is set and when the exposure is done, the mean value of the frame is used to determine the next exposure time. 
The sequence automatically stops when the exposure time gets too long (or too short), but it can also be stopped by the observer at any time. 
To avoid systematic errors from stars in the field, the telescope is moved slightly (dithered) between each exposure, and the observer is also able to reject any bad frames when the sequence has ended. 

When the $k$ exposures have been made, each frame is bias subtracted and the mean value for each $(i,j)$ pixel over all $k$ frames is found. The resulting frame is normalised and used as the master flat-field frame $F$.
Based on an analysis of dome flats made on consecutive nights, we find that the mean variation in pixel value between the master dome flats is $0.0018 \pm 0.0012$.

When using the conventional imaging mode, another readout amplifier is used, and another set of calibration data is therefore needed. 
%No routines for this has been made, but it can be made manually using the \texttt{Conventional} function, that is described below.

\subsubsection{Data acquisition}
There are three ways of acquiring data. The simplest one is the \texttt{snapshot} function, which makes a single integration of the chip, and then reads it out using the EM readout register with the chosen EM gain. The image is then saved directly as a FITS file, without doing any calibration.

To do high frame-rate imaging, one has to use the \texttt{Spool} function. This function acquires the requested number of exposures using the chosen exposure time and EM gain. 
Using the \texttt{PyTables} project \citep{Alted2002}, all of the exposures are saved into a HDF5 format file \citep{hdf5}, along with the necessary calibration frames and header information.
% (A description of the HDF5 files and their content is given in Sect. \ref{sec:hdf5}).
The exposures are acquired from the camera in blocks of 100 exposures, and each block of exposures is then saved to the HDF5 file while the next block is acquired. 
This is done to avoid having to store many GBs of data in the virtual memory of the computer until the acquisition is done, which might lead to an unstable system.
The start of each block gets a timestamp, which is also saved in the HDF5 file together with the necessary metadata for the observation. 
While data are being saved in the HDF5 file, a \texttt{.tmp} suffix is appended to the file name. When the the observation is done and the last exposures are saved, the suffix is removed. 

Parallel with the acquisition a drift correction routine is running. 
The first block of exposures is stacked into a single frame which is stored as a reference. 
Each subsequent block is also stacked and the pixel shift between this frame and the reference frame is found using cross-correlation. 
If the shift is above five pixels, a message is sent to the Odin server, which can then correct the offset. To avoid destroying an observation due to an error in the cross-correlation, an upper correction limit of 10 pixels is set.

A function, \texttt{Conventional}, to do conventional imaging has also been implemented. 
This uses the conventional readout register that has much lower readout noise than the EM register (see Sect. \ref{sec:cam}), and produces a single FITS file. 

\subsubsection{Pointing correction}
Due to the relatively small FoV, $45\arcsec \times 45\arcsec$ (see also Sect. \ref{sec:cam}), it is important that a target can be revisited with a high pointing accuracy. 
The DK154 has a pointing accuracy of about $8\arcsec$ rms, which corresponds to about 20\% of the width of the frame. 

For that purpose, a pointing correction routine has been implemented. 
The first time a target is visited, the pointing is adjusted manually (or just used as it is) and a 10\,s observation of the target is made. This observation is reduced and saved as a FITS file, and is used as a pointing reference. 
The next time the target is visited, the \texttt{PointingCorrection} function will make a new 10\,s observation and compare this to the pointing reference, finding the pixel shift between the two. 
This shift is then returned to the Odin server, so that the offset can be corrected. 
With this pointing correction routine, we can achieve a pointing accuracy better than $0\farcs5$. 
Accuracy below this limit is hard to achieve as the centroid of the stars starts to be dominated by atmospheric disturbances in this regime.

The PSF width (FWHM) of the 10\,s observation is also found and reported back to the SQL database. This can then be used as a measure for the current seeing.

\subsection{Odin server} \label{sec:odinServer}
The Odin server is the top-level module for the Odin system, and links the (virtual) observer, the camera servers, and the TCS. 

When the Odin server is running, it can be connected to one or both cameras, and one can choose to use one or both cameras. 
All the functionality described in Sect. \ref{sec:camServer} is now available. The functions are, however, wrapped in code that makes it possible to operate the two cameras simultaneously, and to report any messages back to the observer. 

All communication with the TCS is gathered in a single module, such that this can be easily customised for different telescopes. 
The TCS module provides the necessary functions to move the telescope, adjust the pointing, and change the focus position. 
It also has functions to get the status of all of the parts of the telescope that are needed to do successful observations. 

Before an observation is started, the camera is checked to make sure it is available for observation (i.e. that it is idle, cooled to the specified temperature, etc.). 
Also a check of the telescope is made, to make sure that observations are possible.
If the celestial coordinates of the target are given, the telescope will be moved to that position, and if a pointing file for the target exists, the pointing will be corrected. 
The observation can now be started, using one or both cameras, depending on which camera is connected and chosen. 
If any problem arises before, during, or after an observation, a message and/or question will be relayed to the observer. 

Provided a list of targets with celestial coordinates and length of the observations (i.e. exposure time $\times$ number of exposures), the Odin server can automatically observe each target on the list, without the need for observer interaction.

\subsubsection{Focus routine}
In addition to the focus system detailed in Sect. \ref{sec:focus}, a routine for focusing the telescope using the EMCCD cameras has been implemented. 
This works by acquiring a number of 10\,s observations at different focus positions and measuring the PSF width of these.
A quadratic fit of the PSF width versus focus position is made, and the minumum PSF width provides the best focus setting. 

It should be noted that the PSF width measurement does not work if the telescope is so much out of focus that donut-shaped PSFs are created. The observer therefore needs to provide an initial estimate of the focus, the focus position step size, and the number of steps. 

After the focus sequence is done, the observer will be able to examine the fit and each focus observation.

\subsection{Odin client}
\begin{figure*}
   \centering
   \includegraphics[width=\linewidth]{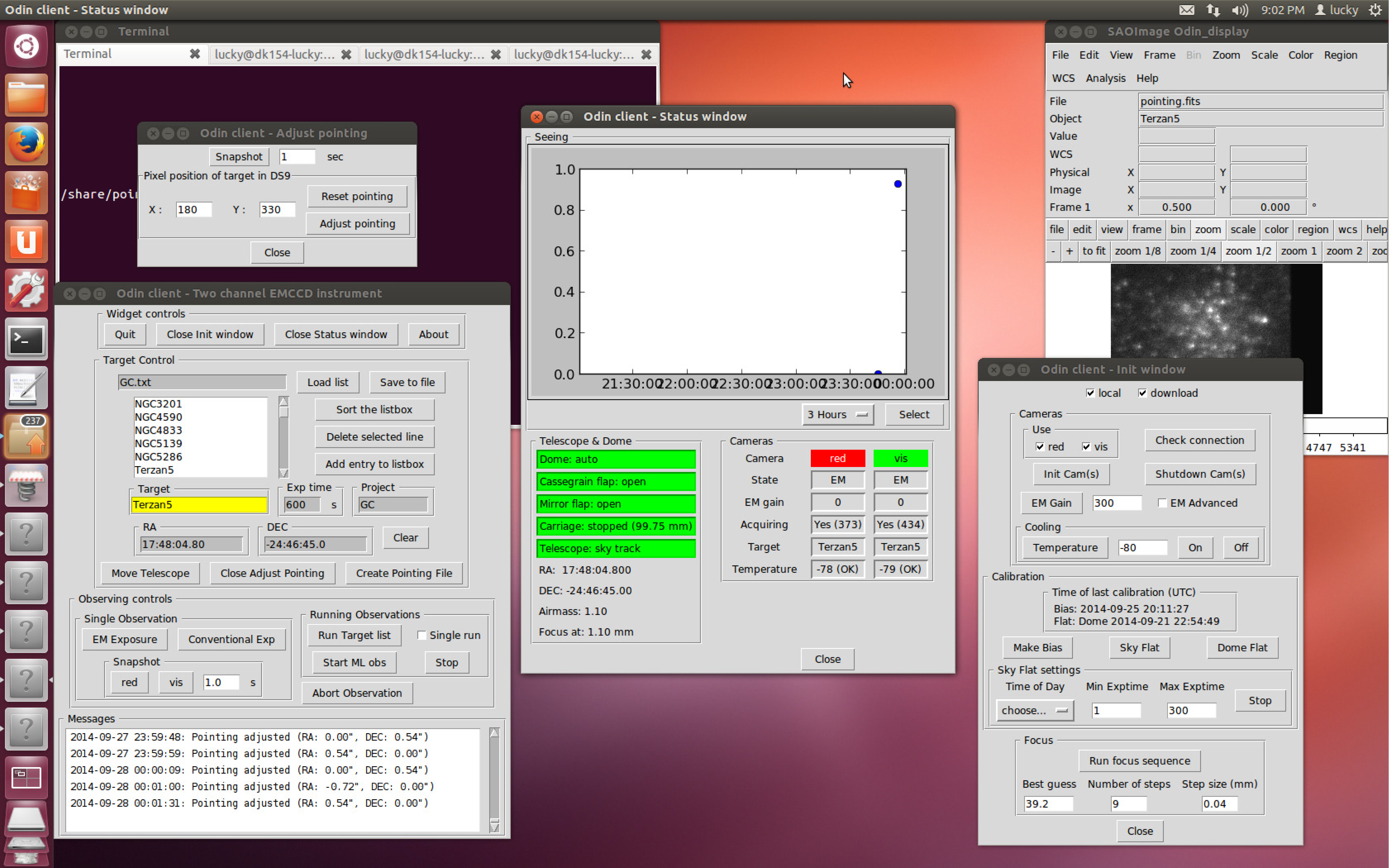}
      \caption{Screenshot of the windows for the Odin client. The main \emph{Odin client} window is shown in the lower left corner. Next to this is the \emph{Status} window, and then the \emph{Init} window below a \emph{SAOimage DS9} window. Above the main window is the \emph{Adjust pointing} window.}
         \label{fig:odin_client}
   \end{figure*}

The Odin client is a GUI for controlling the Odin server. A screenshot of the Odin client and its windows is shown in Fig. \ref{fig:odin_client}.

When starting the Odin client, the main \emph{Odin client} window and an instance of the \emph{SAOimage DS9} FITS file viewer\footnote{http://ds9.si.edu/} are opened. 

The main window contains a list box for selecting and providing the information needed to make an observation. For each entry in the list box, one can give a name, celestial coordinates, the length (in seconds) of the observation, and a project name.
This information can also be read in as a text file. 
The different types of observations can be started, including \emph{Running Observations} that will observe all targets in the list box. 
At the bottom there is a message window that displays messages from the Odin and camera servers. 

The \emph{DS9} window will automatically display the created bias, focus, reference, snapshot images, etc., when these are made.

In the \emph{Init} window different settings for the camera can be controlled, and calibration data and focus sequences can be obtained. 

The \emph{Status} window displays the PSF width measurements obtained by the pointing correction routine over some selectable time interval. It also shows the current status and different values for the telescope and cameras.

To adjust the pointing of the telescope, an $(x,y)$ pixel position from the \emph{DS9} window can be put into the \emph{Adjust pointing} window. The required pointing offset to move the $(x,y)$ pixel position to the middle of the frame will be applied. 

The communication between the Odin client and server has been made light-weight by limiting the amount of information sent between the server and client, and having an option to decrease the size of the frames shown in the ds9 window before they are downloaded. This makes it possible to run the Odin client at a remote location. It is, however, also possible to use VNC (Virtual Network Computing) to remotely control the Odin client. 
%for simplicity a VNC (Virtual Network Computing) server has also been set up, such that the Odin client can be run locally, but controlled remotely. 

\subsection{EM reduction} \label{sec:EMsoftware}
After the HDF5 files have been finalised, they are automatically picked up by the EM reduction software, where each exposure is bias, flat and tip-tilt corrected, and the instantaneous image quality is found, using the algorithm described by \cite{Harpsoe2012}.

As the bias level is subject to variations between the single exposures, the bias correction of an exposure is a two step process. First the bias frame $B$ is subtracted from the exposure. This will correct the overall bias pattern in the frame. Then the offset between the overscan region of the exposure and the overscan region of $B$ is found and subtracted from the exposure. This offset represents the exposure-to-exposure variation in the bias level. 

A flat-field correction is done as in conventional imaging, i.e. by dividing each exposure with the flat-field frame $F$.

The tip-tilt correction and image quality are found using the cross correlation theorem. 
Given a set of $k$ exposures, a comparison image $C(i,j)$ is constructed by taking the average of 100 randomly chosen exposures. 
The cross correlation $P_k(i,j)$ between $C$ and a bias- and flat-corrected exposure $I_k(i,j)$ can be found using 
\begin{equation}
P_k(i,j) = \left| FFT^{-1}\left[FFT(C)\cdot \overline{FFT(I_k)}\right] \right|
\end{equation}
where $FFT$ is the fast Fourier transform.
The appropriate shift, that will correct the tip-tilt error, can now be found by locating the $(i,j)$ position of the global maximum in $P_k$.

A measure for the image quality $q_k$ is found by scaling the maximum value of $P_k$ with the sum of its surrounding pixels within a radius $r$ 
\begin{equation}
q_k = \frac{P_k(i_{\text{max}},j_{\text{max}})}{\sum_{\substack{\left|(i-i_{\text{max}},j-j_{\text{max}})\right| < r \\ (i,j)\neq(i_{\text{max}},j_{\text{max}})}} P_k(i,j)}\ .
\end{equation}
Using this factor, instead of just the maximum value of the pixel values in the frame \citep{Smith2009}, helps to mitigate the effects of fluctuations in atmospheric extinction and scintillation that can happen over longer time scales.

To improve the tip-tilt correction and image quality, a new comparison image $C$ is made based on the 100 exposures with the highest $q$ values. 
The final shifts and $q$ values are then found, using the algorithms described above, based on this new comparison image. 
This is done to avoid the effects of sub-quality exposures among the randomly chosen exposures for the first comparison image.

The FFT transformation is done using the \texttt{FFTW3} library \citep{FFTW05}, linked to Python via the \texttt{PyFFTW} project. This has proven to be about 20 times faster than the FFT implementation from \texttt{NumPy}. 

Cosmic rays from the exposures are detected and corrected using a routine based on algorithms from \citet{Harpsoe2011}. 
As the noise in EMCCD data is exponentially distributed, and not normally distributed as in conventional CCD data, a sigma clipping method cannot be used.
Instead the rate of photons is estimated for each $(i,j)$ pixel in the frame over all $k$ exposures. 
Using this photon rate at each pixel position, the probability $p$ for the observed pixel values in each exposure can be calculated.
If a pixel value is too improbable, a new value is generated from a random number generator based on $p$ and the Erlang PDF.

The reduced exposures are not saved in the HDF5 file, but the necessary information (bias offset, quality factor, pixel shift, and cosmic ray information) is saved for each exposure.
% as described in Sect. \ref{sec:hdf5}. 
Using this information the exposures can easily be reduced, combined, and saved in whatever way is required.
%The default output is the ten-layer FITS cube described in Sect. \ref{sec:fitscube}.

The default output is a ten-layer FITS cube, where each layer represents the sum of some percentage of the shifted-and-added exposures after the exposures have been organised into ascending order by image quality. 
To preserve as much spatial information as possible, the layers have the following percentage cuts: (1,~2,~5,~10,~20,~50,~90,~98,~99,~100). The layers are non-cumulative, however, which means that any exposure is only included in one layer, and the percentage of exposures in the layers are thus: (1,~1,~3,~5,~10,~30,~40,~8,~1,~1).
This means that the first layer contains the sum of the best 1\% of the exposures in terms of image quality, the second layer the second best 1\%, the third the next 3\%, and so on.

\section{System performance} \label{sec:Performance}
This section contains a brief analysis of the improvement in spatial resolution compared to conventional imaging that is achieved with the \tcI, and how updated photometric methods for high frame-rate imaging can improve the photometric precision.

To be able to perform the tip-tilt correction and quality estimates of the individual exposures as described in Sect. \ref{sec:EMsoftware}, it is necessary to have a signal of a certain strength. 
The standard setting for the EM system at the DK154, is therefore to use an EM gain of $\gamma = 300\,\elec/\text{photon}$ and a 10\,Hz frame-rate.
Simulations of stellar populations based on the Besan\c{c}on model \citep{Robin2003}, show that with these settings there will be stars bright enough in, for instance, all pointings towards Baades window to estimate the exposure quality.
Using a higher frame-rate would decrease the signal in the individual exposures, and thus limit the possible pointings. 

\subsection{Spatial resolution} \label{sec:spatial_res}
\begin{figure}
	   \centering
	   \includegraphics[width=\linewidth]{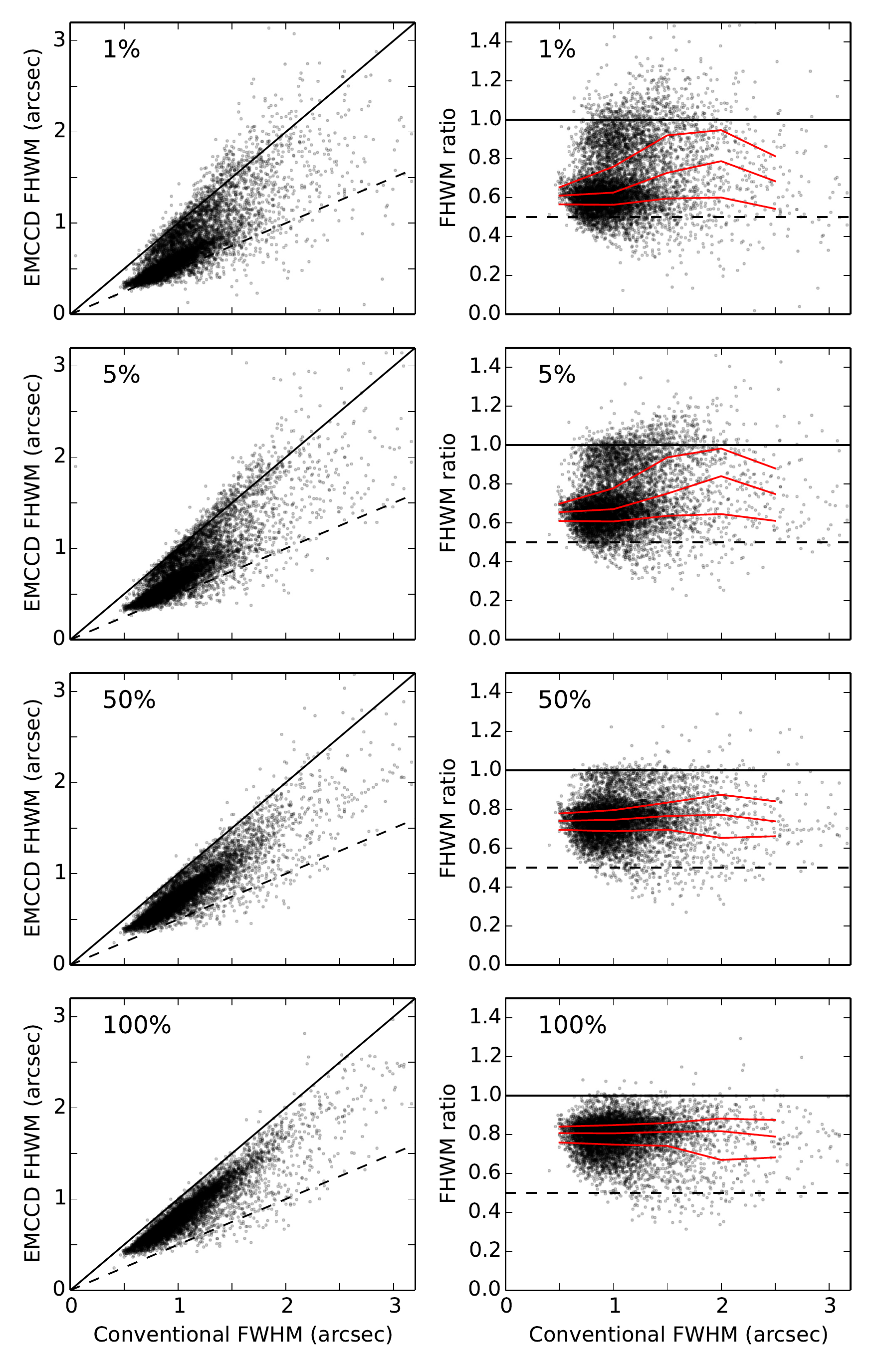}
	      \caption{Left column: For each observation, the FWHM of the 1\%, 5\%, 50\%, and 100\% of the best frames in an observation is compared to the FWHM of the corresponding 'conventional' observation (see text for details). The solid line is unity, i.e. no improvement, while the dashed line shows the half of the conventional FWHM.
	      Right column: Similar to the left column, but showing the ratio between the shift-and-added and 'conventional' FWHM. The red lines give the median (50\%) and the inter-quartile points (25\%, 75\%) of each half-arc second bin.
	      The plots are based on FWHM measurements of 7648 observations from the 2014 season at the DK154.}
	         \label{fig:fwhm}
	 \end{figure}

To get an idea of how the EM system performs in terms of improving the spatial resolution compared to conventional imaging, we have performed a test similar to the one made in \cite{Smith2009}, which is described in Sect. \ref{sec:EMCCD}.
However, instead of looking at Strehl ratios, the improvement in PSF width has been examined.
For over 80\% of the red-band observations from the 2014 season, the FWHM of different percentage selections %of the best exposures in each observation have been shift-and-added into a single image, and the FWHM of each of these images 
have been found.
To mimic an observation made with a conventional CCD, all the exposures in each observation have been stacked without doing any shifts, and from this we find a 'conventional' FWHM.
In the left column of Fig. \ref{fig:fwhm} the 1, 5, 50, and 100 percentage FWHM are plotted against the conventional FWHM. 

We find that $\sim 25\%$ of the FWHMs in the 1\% selection are below $0\farcs5$, while about 8\% are below $0\farcs4$.
For the 100\% selection, these numbers are 4\% and 0.1\%, respectively, while for the conventional FWHM they are 0.1\% and 0.01\%, respectively.
At spatial resolutions below $0\farcs5$, we start to see effects from the telescope optics at the DK154. 
As the telescope is not designed for such high-resolution instruments, a triangular coma originating from the mirror support dominates the PSF more and more as the PSF width decreases.
This effectively puts a limit on the spatial resolution that is higher than the diffraction limit of $\sim 0\farcs13$ at 800\,nm.

To see if a good FWHM for the 1\% selection image also means a good FWHM for the other selections, and vice versa, the correlation coefficient $\rho_{X,Y}$ between the different selections was found.
Between the 1\% and 5\% selection FWHMs a value of $\rho_{1\%,5\%} = 0.94$ was found and this number decreases for larger selections ending in $\rho_{1\%,100\%} = 0.85$.
The same pattern is seen for the other percentages, though with the lowest $\rho$ being $\rho_{5\%,100\%} = 0.91$ and the highest $\rho_{90\%,100\%} = 0.996$. 
Thus there seems to be a fairly good correlation between the FWHMs of the different selections, with the 1\% selection as a slight outlier.

The right column of Fig. \ref{fig:fwhm} shows the improvement in PSF width, as the ratio of the shifted-and-added FWHM to the conventional FWHM. On top of this the median (50\%) and the inter-quartile points (25\%, 75\%) of each half-arcsecond bin is plotted.
%ean and standard deviation for the quartiles of each half-arcsecond bin is plotted.

For the 1\% selection, $\sim 10\%$ of the observations have a FWHM ratio better than 0.5, and the majority are below 0.7, but $\sim 25\%$ of the observations are worse than 0.8.
%For the 1\% selection, the main part of the observations have an FWHM ratio better than 0.7, and about 10\% are below 0.5. For about a fourth of the observations are worse. 
Most of these are between 0.8 and 1, but $\sim 6\%$ are above 1.
This pattern repeats itself for the 5\% selection. 

One reason for this could be that when a short observation is made, only very few exposures go into the 1\% (and 5\%) image. This means that the S/N is very low and this might make it hard to determine the FWHM in a robust way.
A proof of this would be if there were a correlation between the observation length and the FWHM ratio, but this is not found to be the case.
Another reason could be that these bad observations were made at high airmass, but only a very weak correlation of $\rho\simeq 0.4$ has been found between airmass and FWHM ratio, and this value is the same for all selections. 
Another explanation could be that the 10 Hz frame-rate used for these observations is not high enough in some weather conditions, for instance during high winds. 
%This would also explain why $~sim 25\%$ of the observations in the lower percentages selection seems to 
A more thorough analysis of the affected observations and reduction algorithms is needed to sort out this issue, but this is not within the scope of this article.

	\begin{table}
	 \caption{Robust mean and rms values of the FWHM ratios.}
	 \label{table:fwhmImpr}
	 \centering
	 \begin{tabular}{rcccccc}
	 \hline\hline
	 \% & 1 & 5 & 10 & 50 & 90 & 100 \\
	 \hline
	 $\mu$ & 0.64 & 0.69 & 0.71 & 0.75 & 0.79 & 0.81 \\
	 rms & 0.17 & 0.16 & 0.16 & 0.1 & 0.09 & 0.08 \\
	 %$\sigma$ & 0.17 & 0.16 & 0.16 & 0.1 & 0.09 & 0.08 \\
	 \hline
	 \end{tabular}
	\tablefoot{Robust mean ($\mu$) and rms values \citep{Beers1990} of the FWHM ratios between different selection percentages and a 'conventional' observation.}
	\end{table}

For the 50\% and 100\% selections the scatter in FWHM ratio is reduced.
Fig. \ref{fig:fwhm} shows that the FWHM is improved for almost all observations at the higher percentage selections when doing shift-and-add.
Table \ref{table:fwhmImpr} shows the robust mean $\mu$ and rms values \citep[based on median absolute deviation, cf.][]{Beers1990} for the FWHM ratios at different selection percentages. 
This indicates that the average improvement of doing shift-and-add versus conventional imaging is about a fifth of the PSF width on average for the $100\%$ selection, and $\sim 35\%$ for the 1\% selection. 
It also shows that virtually no improvement in FWHM is achieved by omitting the worst 10\% of the exposures. 
%For the lower percentage selections, the mean improvement is clearly highly affected by the 
%The rms values for all selection percentages is around 0.1, which indicates that our method is fairly robust, as the frame-to-frame improvement is more or less the same. 

It should be noted that shifted-and-added EMCCD data generally have a PSF shape that is more like a core-halo structure than a Gaussian profile. 
A FWHM measurement of this kind of profile therefore has an intrinsic uncertainty. 
We have, however, chosen to use this measure, as we believe it provides the best basis for comparison with other systems.

\subsection{Photometric stability}
% Note: the observers were Rabus and Calchi Novati.
The photometric performance of the TCI was investigated in the high-precision domain by monitoring one transit of the planetary system OGLE-TR-56 \citep{2003Natur.421..507K}. This object consists of a gas giant (mass 1.4\Mjup, radius 1.7\Rjup) transiting an F-type star (mass 1.3\Msun, radius 1.7\Rsun) every 1.21\,days \citep{2012MNRAS.426.1291S}. The comparatively large component radii and short orbital period make OGLE-TR-56 a good candidate for the detection of tidally-induced orbital decay \citep{2014MNRAS.440.1470B}. This is potentially detectable by measuring the progressively earlier occurrence of transits compared to a linear ephemeris, by an amount which is predicted to be roughly 30\,s over a decade \citep{2014MNRAS.440.1470B}.

On the night of 2014/05/08 we monitored OGLE-TR-56 (magnitudes $V = 16.6$ and $I=15.30$) for 4.5\,h centred on the predicted midpoint of a transit of duration 2.2\,h. We used the red camera and specified 1000 exposures at a 10 Hz frame-rate with a gain of $300\,\elec/\text{photon}$. The seeing was $\sim 0.5\,\arcsec$ as measured from the best 1\% of images, but the sky conditions were not photometric.

Data reduction was performed on the fits cubes (see Section 6.2) after stacking multiple layers together to obtain images containing the best 90\% of the original observations, corresponding to a combined exposure time of 90\,s. Tests with different numbers of layers showed that the best photometry was obtained using a high fraction of the original images, as expected for this relatively faint star. Differential photometry versus up to five comparison stars was obtained using the {\sc defot} pipeline \citep{Southworth2009} modified to use the PSF photometry routines available in the NASA {\sc astrolib}\footnote{The {\sc astrolib} subroutine library is distributed by NASA. For further details see: {\tt http://idlastro.gsfc.nasa.gov/}.} which are an early version of {\sc daophot} \citep{Stetson1987} ported to the {\sc idl}\footnote{The acronym {\sc idl} stands for Interactive Data Language and is a trademark of ITT Visual Information Solutions. For further details see: {\tt http://www.ittvis.com/ProductServices/IDL.aspx}.} programming language.

\begin{figure} 
		\includegraphics[width=\linewidth,angle=0]{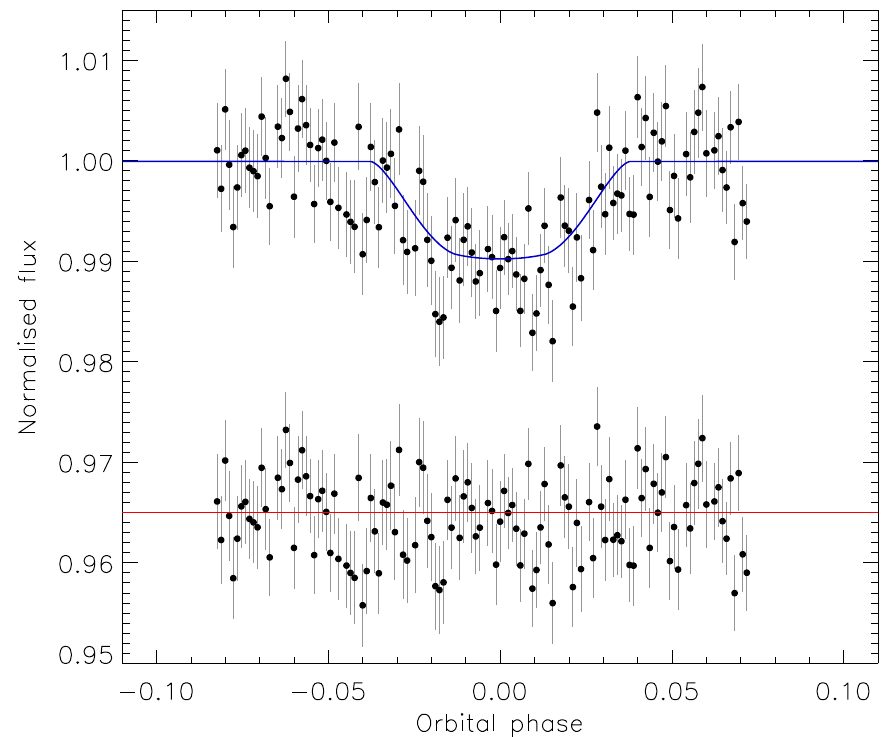} 
		\caption{\label{fig:ogletr56} Light curve of the transiting planetary system OGLE-TR-56 obtained using the red camera of the TCI. 
		The black points show the light curve(top) and residuals (bottom). The blue line shows the {\sc jktebop} best fit and the red line the value the residuals have been shifted to.} 
	\end{figure}

The resulting light curve has a scatter of 4.4\,mmag around the best fit, and shows a clear detection of the transit of depth 10\,mmag (see Fig.\,\ref{fig:ogletr56}). 
For comparison the photometric precision using a conventional CCD and defocussed photometry has been estimated as 3.4\,mmag using the S/N calculator described in \citet{Southworth2009}.
The light curve was modelled using the {\sc jktebop} code \citep{2008MNRAS.386.1644S} with all parameters fixed at known values \citep{2012MNRAS.426.1291S} except for the time of transit midpoint. The transit time was measured to a precision of 314\,s. 
We conclude that the TCI on the Danish telescope is capable of high-precision photometry, but that the results achievable for OGLE-TR-56 are too limited by photon noise to be useful for transit timing studies of this system.

\subsection{Photometric improvement} \label{sec:phot_impr}
Due to the improvement in spatial resolution that the EMCCD cameras can provide, these cameras are optimally suited for crowded fields.

\begin{figure}
   \centering
   \includegraphics[width=\linewidth]{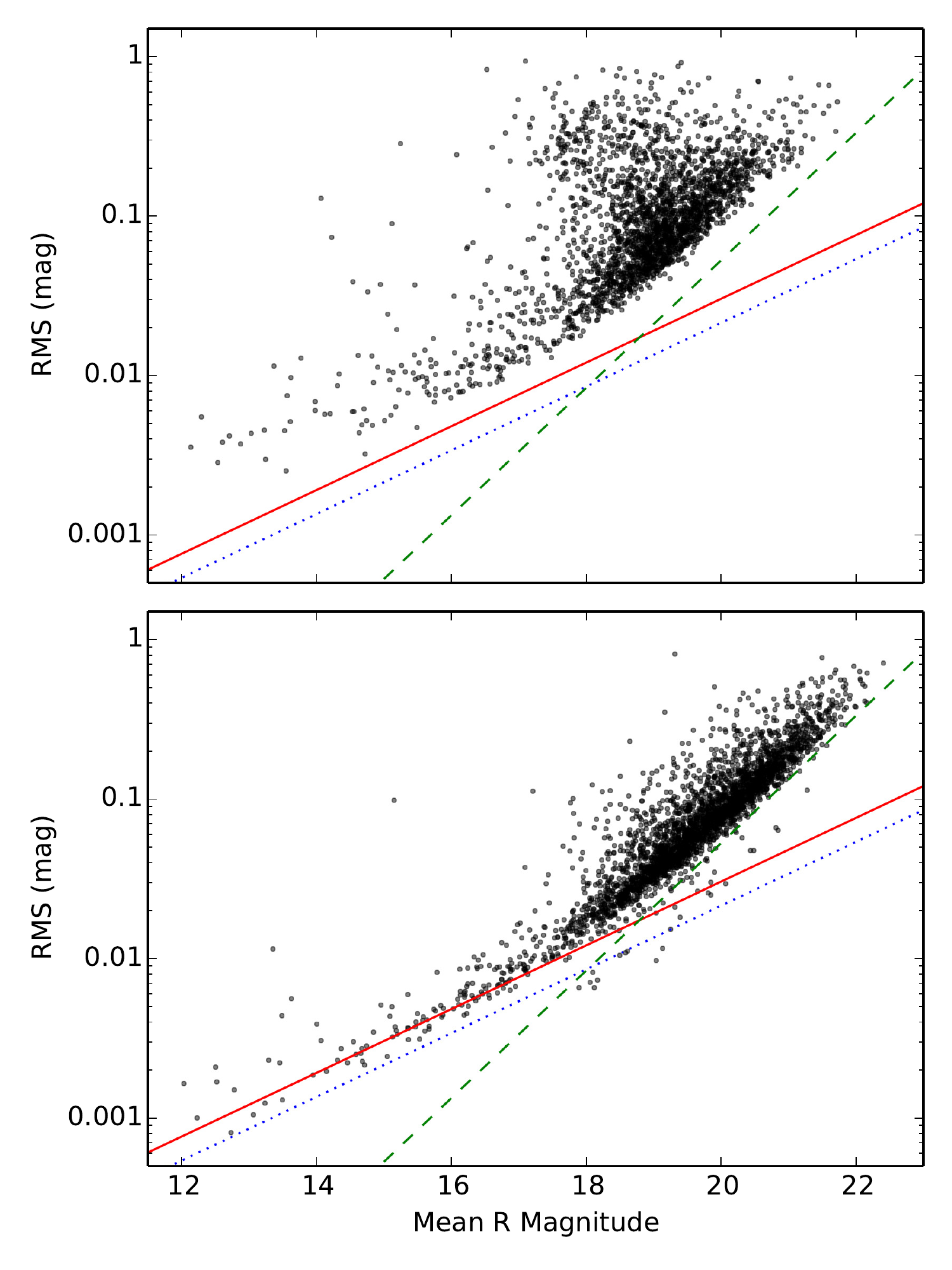}
      \caption{Plots of the RMS magnitude deviation versus the mean magnitude for the 2523 and 3010 calibrated $R$ light curves, extracted by \textsc{DaoPhotII} (upper panel) and the modified version of \texttt{DanDIA} (lower panel), respectively, for the central region of the $\omega$ Cen cluster.
      The photon noise and excess noise limits are shown as a blue dotted, and a red solid line, respectively. The dashed green line shows the noise limit for a 18.5\,mag$/\square\arcsec$ background signal.}
         \label{fig:daodia}
 \end{figure}

To test the photometric stability of the \tci in crowded fields, \citet{Harpsoe2012} have made a photometric reduction of the very dense central part of the globular cluster $\omega$\,Cen. 
The 1.5 hour observation, consisting of $50,000$ single exposures, was stacked into 100 images consisting of 500 consecutive exposures, using the algorithms described in Sect. \ref{sec:EMsoftware}. 
A reference image using the best 1000 exposures was created and this had a FWHM of $\sim 0\farcs4$. 
Using this reference frame the time series photometry was extracted from the 100 images using the standard PSF fitting photometry package \textsc{DaoPhotII} \citep{Stetson1987}.
The upper panel of Fig. \ref{fig:daodia} shows the RMS magnitude deviation for the 2523 light curves extracted by \textsc{DaoPhotII}. 
The light curves have been approximately calibrated to Johnson $R$ magnitudes using aperture photometry of 3 reasonably isolated stars, and their corresponding F675W filter magnitudes taken from Hubble Space Telescope WFPC2 camera data.
The plot also shows the photon noise and excess noise limits, and the noise limit for a 18.5\,mag$/\square\arcsec$ background signal. 
%It is evident that the \textsc{DaoPhotII} reduction was off on the limits. 

To see if the photometric reduction could be improved, the same $\omega$ Cen dataset was reduced using the \texttt{DanDIA}\footnote{{\tt DanDIA} is built from the DanIDL library of IDL routines available at http://www.danidl.co.uk} pipeline \citep{Bramich2008, Bramich2013}.
\texttt{DanDIA} uses difference imaging analysis (DIA), which is particularly apt at extracting precise photometry in crowded regions.
Instead of modelling the kernel as a combination of Gaussian basis functions, \texttt{DanDIA} uses a flexible discrete-pixel kernel. 
This has been shown to give improved precision in very crowded regions \citep{Albrow2009}. 

To accommodate the advantages of high frame-rate imaging, a few modifications have been made to the pipeline. 
For each observation, the single exposures can be combined in two ways:
	\begin{description}
	\item{Quality-binned:} exposures are binned after they have been organised into ascending order by image quality.
	\item{Time-binned:} exposures are grouped into time bins, such that the required S/N is achieved. 
	\end{description}
The quality- and time-binned images can either be constructed directly from the HDF5 spools, as described in \citet{Skottfelt2013}, or they can be made from the FITS cubes. If the latter method is used \citep[see][]{Skottfelt2014}, the single layers can be saved as quality-binned images, and the whole cube, or some subset hereof, can be stacked into a time-binned image.
The pipeline is then able to stack the sharpest of the quality-binned images to make a high-resolution reference image from which reference fluxes and positions are measured. 
The reference image, convolved with the kernel solution, is subtracted from each of the time-binned images to create difference images, and, in each difference image, the differential flux for each star is measured by scaling the PSF at the position of the star \citep[cf.][]{Bramich2011}.
Compared to conventional CCD data, EMCCD data has different noise model \citep{Harpsoe2012}, and this has also been implemented. 

The RMS magnitude deviation for the 3010 light curves extracted with \texttt{DanDIA} is shown in the lower panel of Fig. \ref{fig:daodia}. Johnson $R$ magnitudes have been approximated using the same three stars as in the \textsc{DaoPhotII} reduction.
It is apparent that the DIA reduction is superior to the \textsc{DaoPhotII} reduction, in that the majority of the photometric scatter falls much closer to the theoretical limits. 
There is an improvement of the mean rms for all magnitude bins, ranging from a factor of $\sim 2$ for the brightest magnitudes, to a factor of 7 at 19th mag.
For magnitudes fainter than about 19, the photometric precision is limited by the equivalent of a 18.5\,mag$/\square\arcsec$ background signal, which most likely originates in the unresolved dense stellar population in the central part of the cluster.

\subsection{Two-colour capabilities}
With a red and visual camera running simultaneously, it is easy to obtain colour information for the observed stars. 
As the visual camera only came online this summer (2014), no scientific results have yet been published using visual band data from the \tcI. 

To test the two-colour capabilities of the \tcI, six observations of the central parts of the globular cluster NGC~288 (\RA{00}{52}{45.24}, \Dec{-26}{34}{57.4} at J2000.0) were obtained on the night of 21 September, 2014. The observations were done in the red and visual band simultaneously, and the data were analysed using the modified version of the \texttt{DanDIA} pipeline described above. 
The instrumental magnitudes found by the pipeline are denoted $v$ and $i$ for the visual and red band, respectively, and are based on reference images combined from 6780 and 6060 exposures, respectively, which corresponds to $\sim 10$ minutes of observation.
A PSF width of $0\farcs48$ and $0\farcs46$ FWHM are found for the $v$ and $i$ reference images, respectively.

\begin{figure*}[t]
   \centering
   \subfloat[][Visual reference image]{\centering
                \includegraphics[width=0.32\linewidth,angle=270]{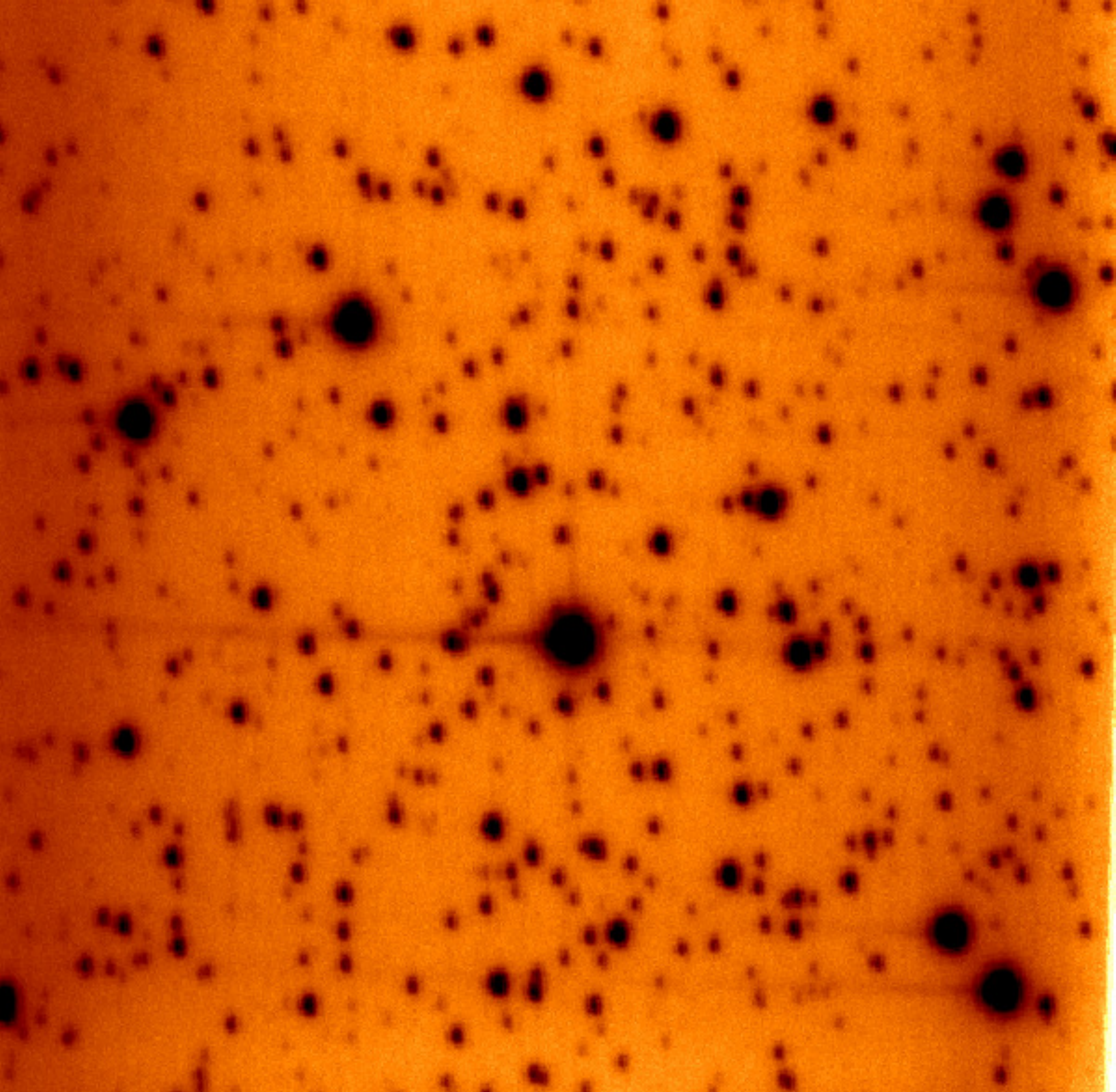}
                \label{fig:NGC288_V}} \hspace{-0.5pt}
   \subfloat[][Red reference image]{\centering
                \includegraphics[width=0.32\linewidth,angle=270]{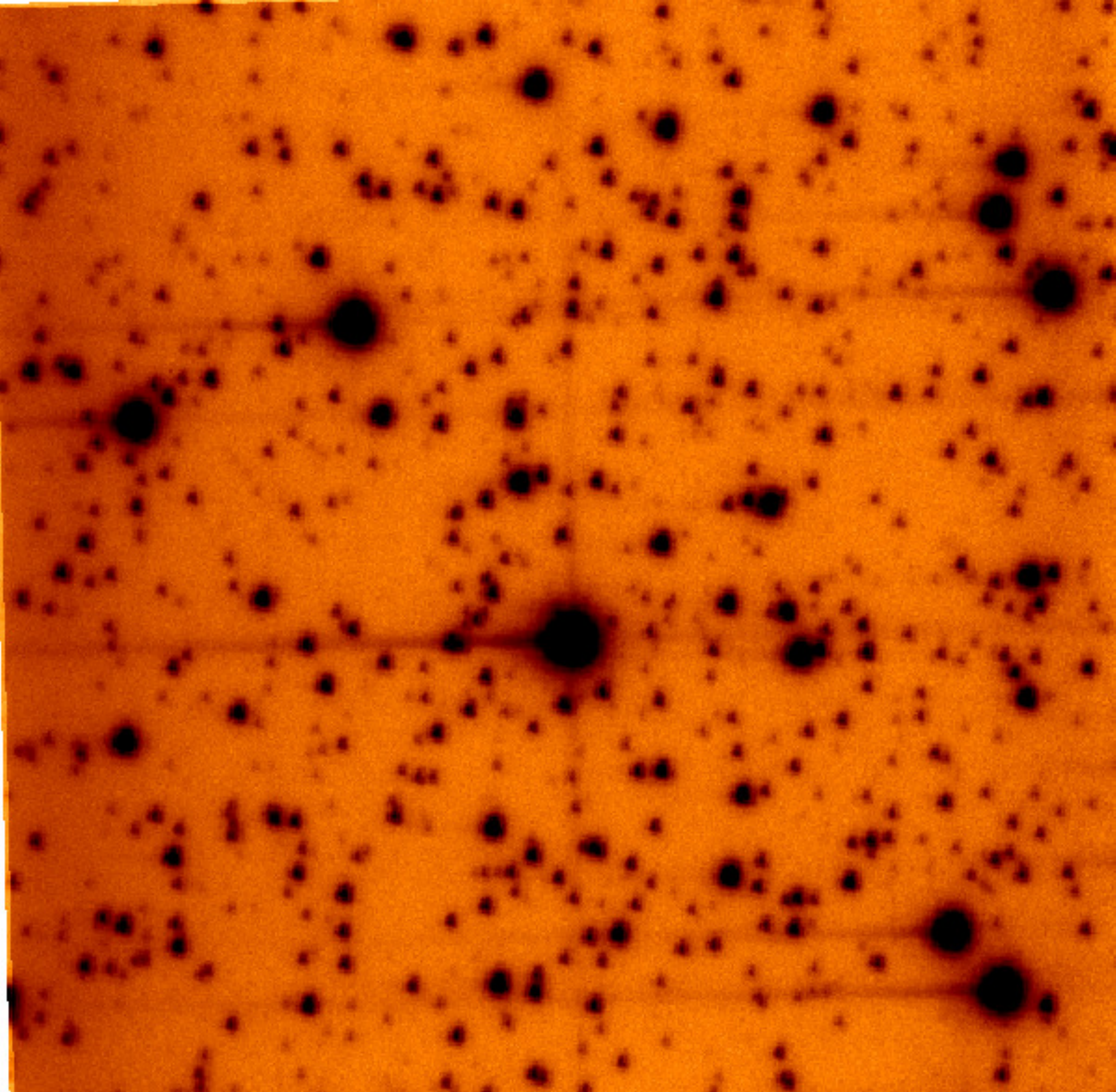}
                \label{fig:NGC288_I}} \hspace{0.5pt}
   \subfloat[][HST ACS image]{\centering
                \includegraphics[width=0.32\linewidth,angle=270]{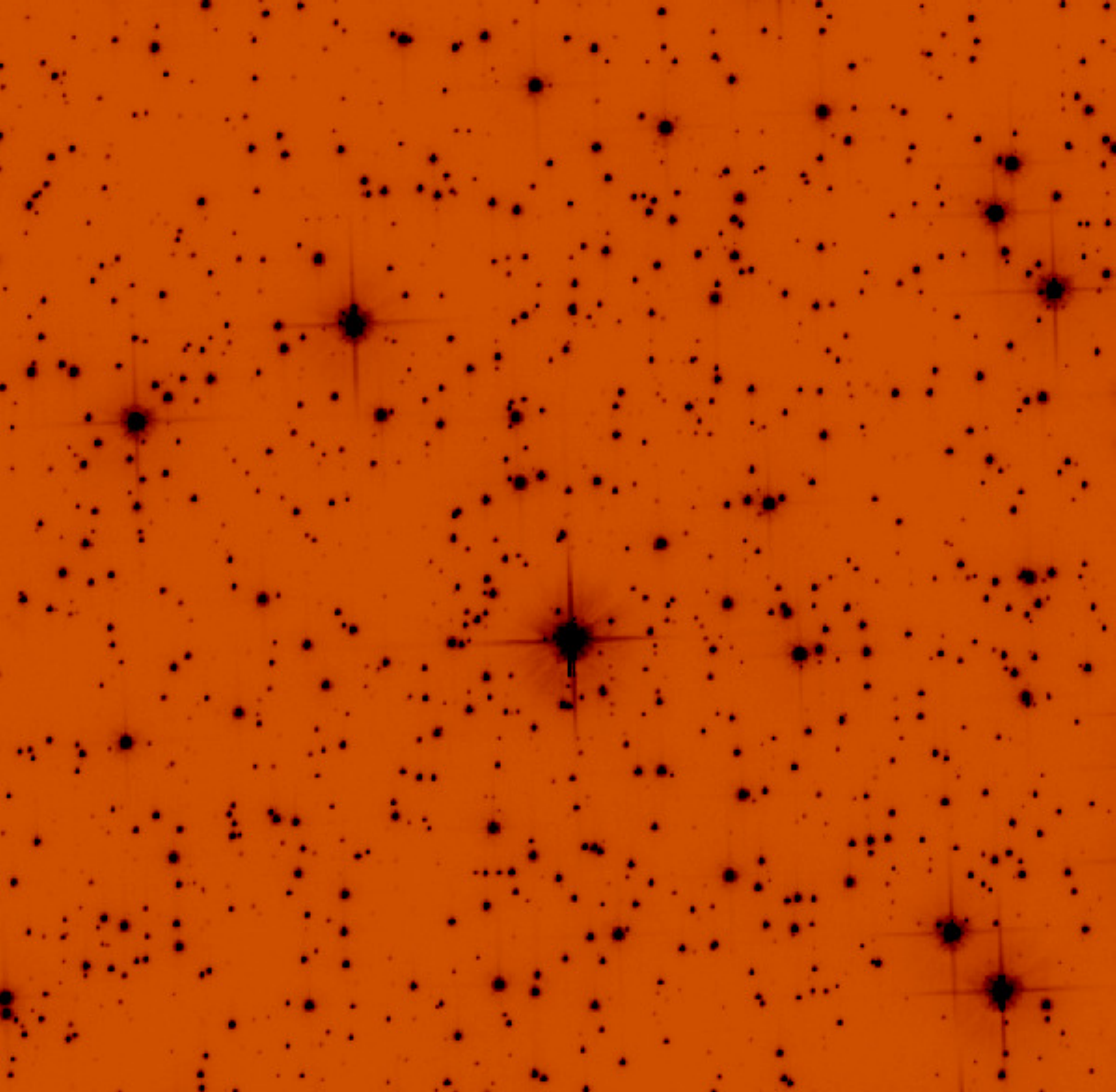}
                \label{fig:NGC288_HST}} %\hspace{0.5pt}   
      \caption{Visual $v$ and red $i$ reference frames for the NGC~288 cluster and a HST ACS image of the same field as comparison. North is up and east is to the left. The image size is about $45\arcsec \times 45 \arcsec$. Note that the flux scale is logarithmic. }
        \label{fig:NGC288_ref}
   \end{figure*}

The NGC~288 cluster is part of \emph{The ACS Survey of Galactic Globular Clusters} \citep{Sarajedini2007} made with the ACS instrument on the Hubble Space Telescope (HST), and the survey is available online\footnote{www.astro.ufl.edu/$\sim$ata/public\_hstgc/}.
The data files include celestial coordinates, ACS F606W filter Vega magnitudes (denoted $V$), and ACS F814W filter Vega magnitudes (denoted $I$) \citep{Sirianni2005}, 
Fig. \ref{fig:NGC288_ref} shows the visual and red reference images for NGC~288 together with a HST ACS image of the same field.
Because the \tci reference images are shown with logarithmic flux scale, two overlapping effects are visible for the brightest stars; a faint cross caused by the mount (or \textit{spider}) of the secondary mirror, and a slightly brighter trail caused by Charge-Transfer Inefficiency (CTI) in the CCD chip. The effect of these is, however, negligible and has not been found to limit the photometric precision of bright stars.

\begin{figure}
	   \centering
	   \includegraphics[width=\linewidth]{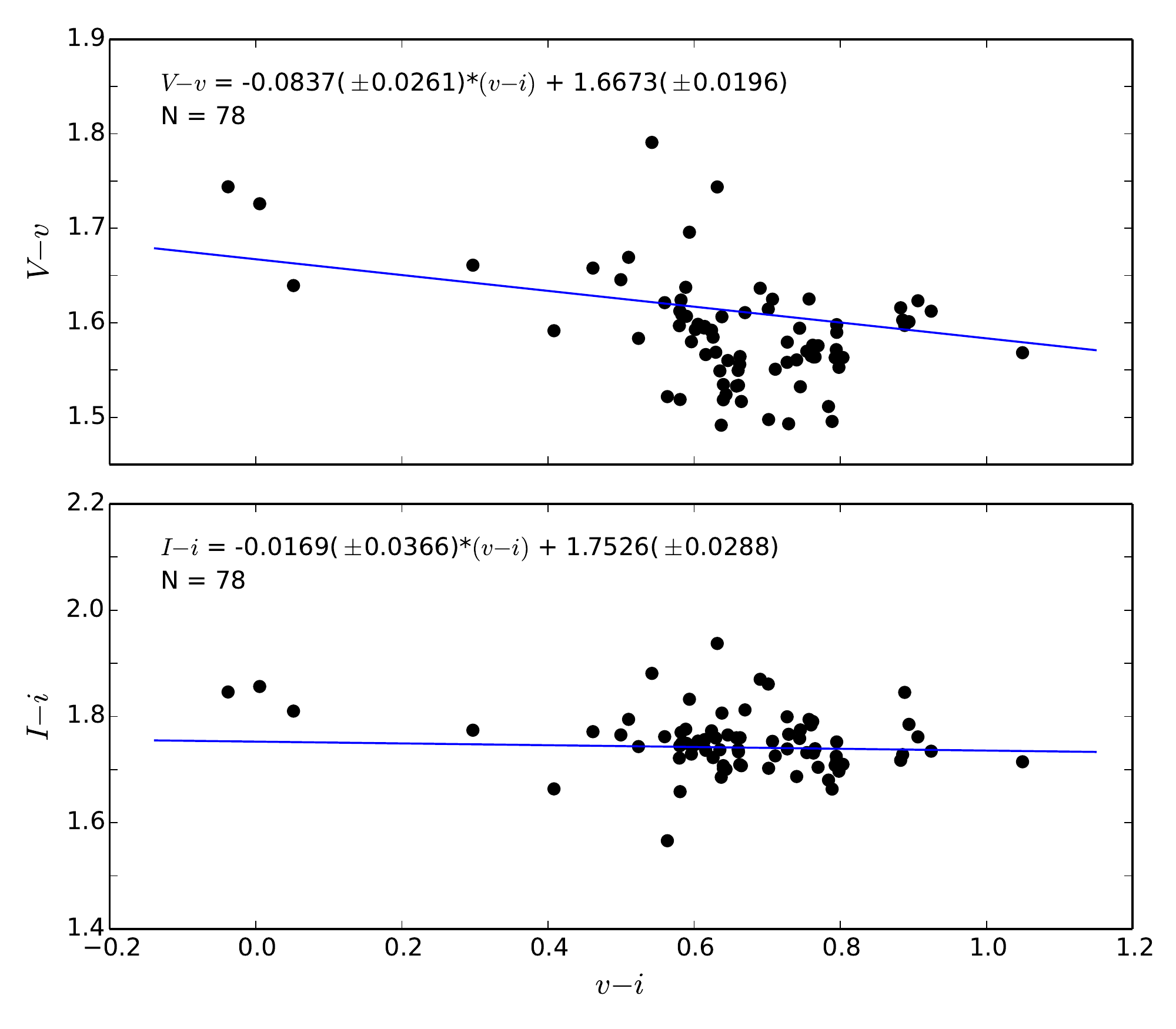}
	      \caption{Relations used to convert from instrumental to ACS magnitudes for the $V$ (top) and $I$ (bottom) bands.}
	         \label{fig:phot_cal}
	 \end{figure}

Using the celestial coordinates we were able to find 430 matching stars, and from these a conversion between the instrumental magnitudes and ACS magnitudes was found. This was done by fitting a linear relation to the difference between the ACS and instrumental magnitudes, and the instrumental $v-i$ colour of each of the stars with $V < 19$. 
The resulting relations are shown in Fig. \ref{fig:phot_cal}.
%We found that if stars fainter than $V = 19$ was used in the fit, more outlier points was introduced, which made the fit unstable. 

	 \begin{figure}
	   \centering
	   \includegraphics[width=\linewidth]{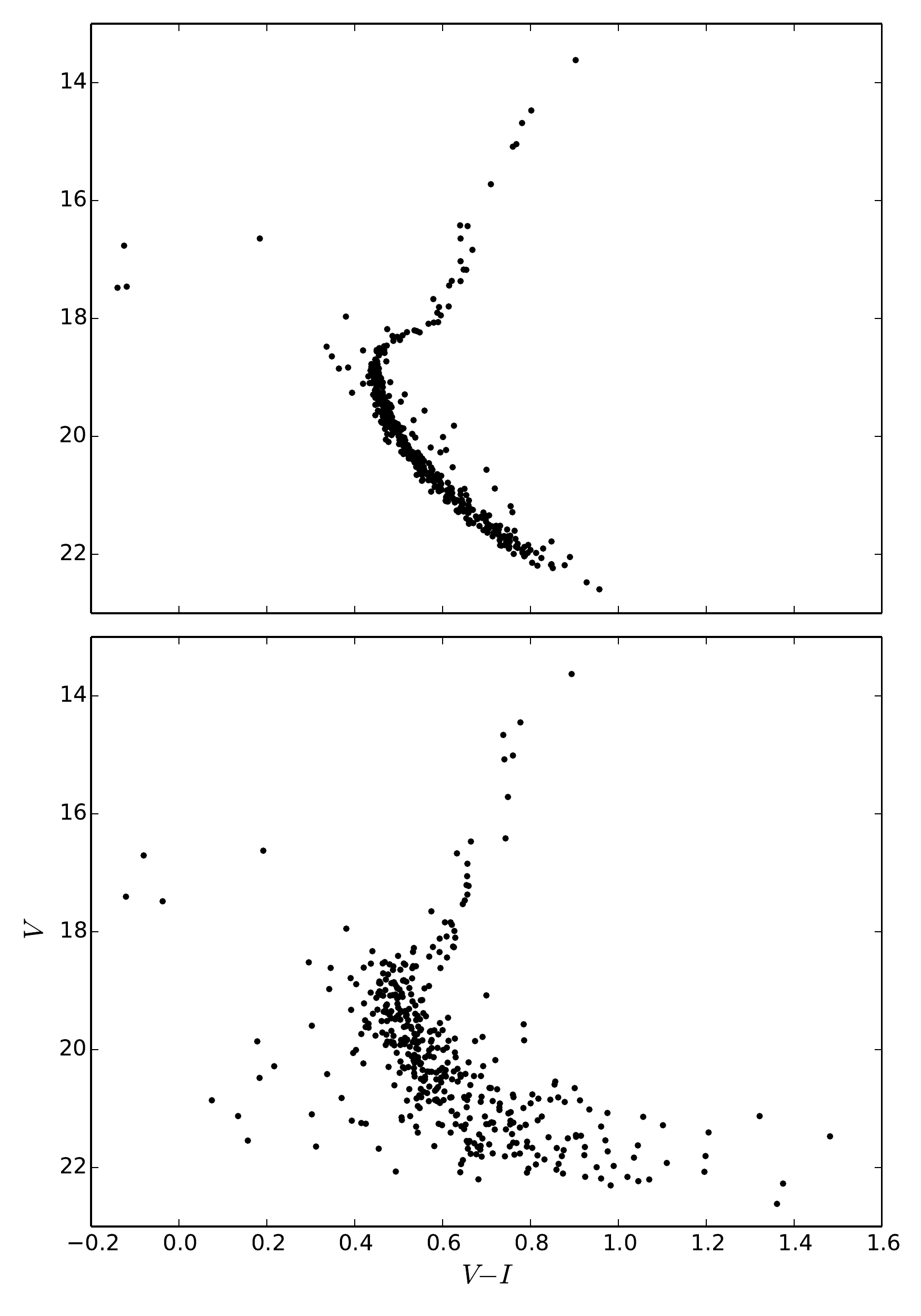}
	      \caption{$(V-I),V$ Colour-magnitude diagram of 430 stars in the central part of NGC~288. Upper panel are the HST/ACS data, and the lower panel are \tci data calibrated using the HST/ACS data.}
	         \label{fig:phot_cmd}
	 \end{figure}

Using the calibrated $V$ and $I$ magnitudes, a colour-magnitude diagram was produced. This is shown in Fig. \ref{fig:phot_cmd} together with the corresponding HST/ACS photometry. 
Especially for the brighter stars there is a good agreement between the two sets of photometry. For the fainters stars there seems to be the same trend, but with a much higher scatter in the \tci data. 
Some differences between the two data sets are not surprising considering that there are difference between the HST/ACS and \tci filters and that we are comparing photometry of faint stars from a 1.5m ground telescope with a 2.5m space telescope.

%However, for stars below $V \sim 20$ mag the scatter becomes very pronounced. 
\citet{Mattila1996} reported the sky brightness for a moonless night at the La Silla observatory as being $V = 21.75(\pm 0.08)\,\text{mag}/\square\arcsec$ and $I = 19.48(\pm 0.05)\,\text{mag}/\square\arcsec$.
The background signal will therefore contribute significantly to the noise at the fainter magnitude levels and leads to increased scatter. 
Signal from background stars will also have an effect on the scatter, but as the central region of NGC~288 is not very dense, the contribution is negligible compared to what was found for $\omega$ Cen. 

%the background signal due to crowding will be somewhat less than what is seen in Fig. \ref{fig:daodia}.
%, where the photometric precision is limited by a background signal corresponding to 18.5 mag$/\square\arcsec$. 

%Also the fact that the ACS survey contains photometry on stars as faint as to 30th mag, means that the some of the stars in the 

\section{Scientific results} \label{sec:Science}
Since 2012 the red EMCCD camera has been used for regular operations at the DK154, initially as the single red camera and during the last two years a full version of the \tcI. 

The main purpose of the \tci is to follow up gravitational microlensing events towards the Galactic centre in order to detect exoplanets, i.e. planets orbiting stars other than the Sun. 
However, a number of other scientific results have also been achieved in this period.

%The first major result from the DK154 that demonstrated the power of EMCCDs for high-precision time series photometry in crowded fields, was the discovery of two previously unknown variable stars near a bright star in the well-studied globular clusters NGC~6981 \citep{Skottfelt2013}.

%Another proof of the EMCCD capabilities was the discovery of two rings around the Centaur (10199) Chariklo. 
%Chariklo is a asteroid-like object orbiting betweeen Saturn and Uranus. With a radius of about 125\,km, it became the smallest known object to have rings. 

%These and other scientific results from the \tci are presented below. 

\subsection{Gravitational microlensing of low-mass exoplanets}
%The main purpose of the \tci is two do follow-up of gravitational microlensing events towards the galactic center, in order to detect exoplanets, i.e. planets orbiting stars other than the Sun. 

During the last 20 years over 2000 exoplanets have been detected using a number of different techniques. 
The by far most successful techniques, in terms of number of detections, are the radial-velocity technique and observations of transiting planets. 
These techniques are, however, skewed towards detecting high-mass planets orbiting close to the host star, and this has left us with an incomplete picture of the exoplanet population statistics \citep{Gaudi2012} %(see Fig.\ref{fig:ML_pop})
%
% \begin{figure}
%   \centering
%   \includegraphics[width=\linewidth]{plot_dao_dia}
%      \caption{Plot of \citep[figure 6 from][]{Gaudi2012}. }
%         \label{fig:daodia}
% \end{figure}

The only technique currently available for obtaining population statistics of cool low-mass planets is gravitational microlensing  \citep{Mao1991}. 
When a background star and a foreground star become near-perfectly aligned with an observer, the foreground star will act as a gravitational lens, magnifying the light from the background star.
Such an event lasts about a month, and creates a characteristic bell-shaped light curve for the background star.
If a planet is orbiting the lens star, it will perturb the gravitational lensing effect, and thus create a feature in the light curve, lasting from days to hours, depending on the mass of the planet.
Subsequent analyses and modelling of the light curve can provide information about the lens star and planet, such as masses and distances, etc.

To detect planets of Earth-mass and below, one needs to have a photometric precision of Galactic bulge main sequence stars to within a few per cent, and this requires angular resolutions below $\sim 0\farcs4$ \citep{Bennett2002}. 
Reaching such high spatial resolution from the ground is difficult with conventional CCD imaging unless an AO system is available, but it can be done with EMCCDs. %, as these can counteract the blurring effects from atmospheric disturbances.
As discussed in Sect. \ref{sec:spatial_res}, the \tci at the DK154 is only just capable of obtaining angular resolutions below $0\farcs4$, due to imperfections in the optics. 
With the near diffraction limited optical system on the SONG telescopes, this technique should be able to achieve spatial resolutions down to $\sim 0\farcs2$ at 800\,nm.
Preliminary test observations during the summer of 2014 reached a resolution of $\sim 0\farcs25$.

Observations made with the \tci at the DK154 are done as part of the MiNDSTEp consortium \citep{Dominik2010}, and this will also be the case for SONG.

\subsection{Variable stars in the cores of globular clusters}
\begin{figure}
		\centering
		\includegraphics[width=\linewidth]{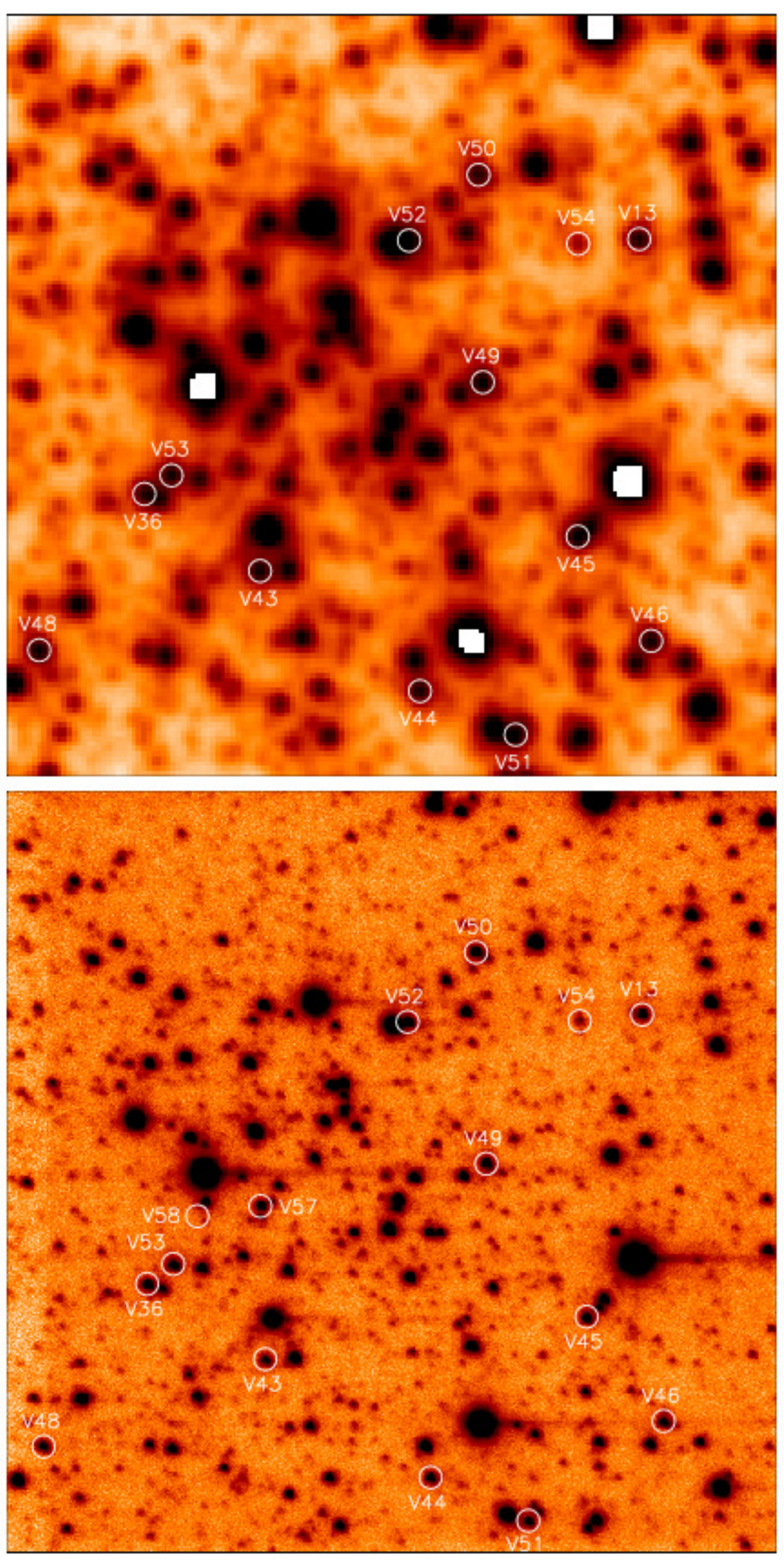}
		\caption{Top: Cut-out of the B11 V reference image corresponding to the field of view of the red EMCCD camera. 
		         Bottom: Reference image from S13, giving the positions of the variables confirmed by B11 and the two new variables V57 and V58 (note that V58 is not the star that is located at the upper right edge of the circle, but a star that is too faint to be detected at the epoch of the reference image).
		         North is up and east is to the right. The image size is $45 \times 45 \text{arcsec}^2$. Notice the greatly improved resolution compared to the B11 finding chart.
		         (Reproduction of figure 1 in S13)}
		\label{fig:NGC6981_ref}
	\end{figure}

Based on observations made with the red EMCCD camera over two nights in August 2012, \citet[][hereafter S13]{Skottfelt2013} were able to find two previously unknown variable stars in the crowded central region of the globular cluster NGC~6981. 
The variable star population of the cluster had been examined previously, latest in \citet[][hereafter B11]{Bramich2011}, where 11 new RR Lyrae (RRL) stars and 3 new SX Phoenicis stars were found.
The two new variable stars, denoted V57 and V58, are located close to a bright star as can be seen in Fig. \ref{fig:NGC6981_ref}. 
In B11, which used conventional CCD imaging, the bright star was saturated and the two variable stars were therefore in a region that could not be measured due to the saturated pixels.
The S13 data was analysed using the modified version of DanDIA described in Sect. \ref{sec:phot_impr}, and from the light curves retrieved, V57 was found to be an RRL star, while it was not possible to classify V58.

Shortly after the publication of S13, another analysis of the variable star population in NGC~6981 was published by \citet{Amigo2013}. 
This analysis was based on archive data from, among other, the FORS1 instrument made at one of the 8.0m VLT telescopes. 
Here the authors were also able find V57, and two other RRL stars outside the FoV of B11, but they were not able to detect V58. 

The results from S13 lead to the start of a survey of variability in Galactic globular clusters using the \tci \citep{Kains2014limbo}. The survey has so far been running at the DK154 during the 2013 and 2014 MiNDSTEp observing campaigns, with bi-weekly observations of about 30 globular clusters.
The first results from the survey have recently been published in \citet[][hereafter S14]{Skottfelt2014}. 

In S14 the analysis of time series observations from 2013 and 2014 of five metal-rich ([Fe/H] $>$ -1) clusters is presented. 
In the analysis 48, 49, 7, 8, and 2 previously unknown variables are found in NGC~6388, NGC~6441, NGC~6528, NGC~6638, and NGC~6652, respectively. 
Due to the relatively small FoV of $45\arcsec \times 45\arcsec$ of the \tcI, only the dense central part of the clusters is observed. 

Especially interesting is the case of NGC~6441, for which the variable star population of about 150 stars has been thoroughly examined by previous studies, including a HST study. 
Of the 49 new variable stars presented in the S14 article, one (possibly two) are RRL stars, two are W Virginis stars, and the rest are long period semi-regular/irregular variables on the red giant branch. 
Furthermore the first double mode RR Lyrae was detected in the cluster. 

S13 and S14 thus demonstrate the power of EMCCDs combined with DIA for high-precision time-series photometry in crowded fields, and the feasibility of large-scale observing campaigns using these methods.

\subsection{The rings of Chariklo}
On June 3rd, 2013, the Centaur (10199) Chariklo occulted a background star.
Chariklo is an asteroid-like object of about 125 km in radius, orbiting between Saturn and Uranus with a semi-major axis of 15.8 AU. 

As reported by \citet{BragaRibas2014} the stellar occultation by Chariklo was observed from eight sites in South America. 
The occultation by Chariklo itself was only recorded at three sites in Chile, but seven sites detected an indication of a secondary event in the form of rapid stellar flux interruptions.
However, it was only from the DK154 data that the ring structure could be clearly identified. 
With a 10 Hz frame-rate, the \tci was able to resolve these secondary events into two sub-events, lasting only  $\sim 0.1$ and $\sim 0.3$\, seconds, with a 0.2 second gap. 

The best interpretation of the observations is that Chariklo has two rings. 
With the high precision data from the \tcI, the width of the two rings was determined to be $\sim 7$ and $\sim 3$\,km wide, with a few hundred metres precision, and located at orbital radii of $\sim 391$ and $\sim 405$\,km, respectively. 

The presence of rings, partly composed of water ice, around Chariklo explains a dimming observed in the Chariklo system between 1997 and 2008 and the gradual disappearance of ice and other absorption features in its spectrum over that period.

Previous stellar occultations by main-belt asteroids have not revealed any rings, so it is not known if rings around minor bodies are generic or exceptional features. 
However, the fact remains that Chariklo is, so far, the smallest solar system object known to have rings.

\subsection{Locating faint stars near transiting planetary systems}
Using defocused observations of transiting planets, it has been possible to reach photometric precisions down to a few tenths of a millimagnitude with the DFOSC instrument at the DK154 \citep[e.g.][]{Southworth2009,Southworth2010}.
There is however a risk that faint nearby stars can contaminate the signal from the target stars. 
Such objects would dilute the transit and cause the radius of the planet to be underestimated, or, if the contaminant is an eclipsing binary, render the planetary nature of the system questionable.

To search for faint stars close to the target star which are not visible on conventional CCD images, the \tci can be used to observe the target stars, as the high frame-rate imaging capabilities of the \tci allows for long exposures of a bright target, without saturating the CCD. 
After the observation, some percentage of the best exposures can be combined to get a high dynamical range without losing spatial resolution. 
In \citet{Southworth2014} and \citet{Mancini2014} this method was used, and it was found that there were either no nearby stars, or that the closest stars could only have a negligible effect on the results.

\section{Conclusion}
A two-colour EMCCD instrument for the SONG 1m telescopes and the Danish 1.54m telescope has been developed and we have detailed the design and implementation of this instrument. 
The \tci is the first routinely operated multi-colour instruments providing LI photometry. 
The Odin software framework enables these routine observations and has been prepared to control the \tci locally and remotely.

The EMCCD cameras are used to do high frame-rate imaging in the visual and red bands simultaneously.
We show that by shifting-and-adding the EMCCD data, the spatial resolution is improved compared with conventional imaging, even at the 10 Hz frame-rate, that is the default at the DK154.

For a 1\% selection we find that the FWHM is 35\% sharper on average compared to conventional observations, and for the best $\sim 10\%$ of the 1\% selections the FWHM is improved by 50\% or more.
For the 100\% selection the FWHM is about 20\% sharper on average.
At the 1\% selection, $\sim 25\%$ of all observations have a FWHM below $0\farcs5$, while $\sim 8\%$ are below $0\farcs4$. 
These numbers drop to 4\% and 0.1\%, respectively, at the 100\% selection, while for conventional imaging they are an order of magnitude lower. 

Imperfections in the DK154 optics introduce a triangular coma that becomes very distinct at spatial resolutions below $0\farcs5$, and thus limit the achievable resolution. With the near diffraction limited system on the SONG telescopes, we expect to reach spatial resolutions down to $0\farcs2$. 

We find that it is possible to reach a photometric precision of a few millimagnitudes for a $I=15.3$\,mag star using the \tci and that high-precision photometry of EMCCD data in crowded fields is possible. 
Using \texttt{DanDIA} (a difference image analysis pipeline using discrete pixel kernels), we are able to achieve a photometric precision several times better than that achieved with the PSF fitting photometry package \Daophot\ for observations of the dense globular cluster $\omega$\,Cen.

EMCCD observations have been performed on a regular basis at the DK154 for several years now. 
This has lead to a number of interesting results including the detection of new microlensing exoplanets, the detection of previously unknown variable stars in the cores of crowded, and well-studied, globular clusters, and the discovery of rings around the asteroid-like object Chariklo, which thus became the smallest solar system object known to have rings.

\begin{acknowledgements}
The operation of the Danish 1.54m telescope is financed by a grant to UGJ from the Danish Natural Science Research Council (FNU). 
We also acknowledge support from the Center of Excellence Centre for Star and Planet Formation (StarPlan) funded by The Danish National Research Foundation.
DMB acknowledges support from NPRP grant \# X-019-1-006 from the Qatar National Research Fund (a member of Qatar Foundation).
MH acknowledges support by the Villum Foundation.
Funding for the Stellar Astrophysics Centre is provided by The Danish National Research Foundation (Grant DNRF106). 
%OW (FNRS research fellow) acknowledges support from the Communaut\'e francaise de Belgique - Actions de recherche concert\'ees - Acad\'emie Wallonie-Europe..

Some of the data presented in this paper were obtained from the Mikulski Archive for Space Telescopes (MAST). STScI is operated by the Association of Universities for Research in Astronomy, Inc., under NASA contract NAS5-26555. Support for MAST for non-HST data is provided by the NASA Office of Space Science via grant NNX13AC07G and by other grants and contracts.
\end{acknowledgements}

\bibliographystyle{aa}
\bibliography{TwoColorEM}   
  
\end{document}